\documentclass[preprint2]{aastex}

\usepackage{epstopdf}
\usepackage{graphics}
\usepackage{lscape}
\usepackage{psfig}
\newcommand{\xrtpos}{\mbox{RA=10$^{\rm h}$58$^{\rm m}$31\farcs13, Dec =$+67\degr$31~$\arcmin$30\farcs8}}

\newcommand{\uvotRA}{\mbox{RA=10$^{\rm h}$58$^{\rm m}$31\farcs12}} 
\newcommand{\uvotDec}{\mbox{Dec =$+67\degr$31~$\arcmin$31\farcs2}}

\newcommand{\xrtposratwo}{\mbox{RA(J2000)=02$^{\rm h}$51$^{\rm m}$51$\fs37$}}
\newcommand{\xrtposdectwo}{\mbox{Dec(J2000)=$+49\degr$16$\arcmin$21\farcs2}}
\newcommand{\uvotRAtwo}{\mbox{RA=02$^{\rm h}$51$^{\rm m}$51\farcs40}} 
\newcommand{\uvotDectwo}{\mbox{Dec =$+49\degr$16~$\arcmin$23\farcs6}}

\newcommand{\NH}{\mbox{${\rm N}_{\rm H}$}} 

\def\CIVdblt{{\rm C~}\kern 0.1em{\sc iv}~$\lambda\lambda 1548, 1550$}
\def\MgIIdblt{{\rm Mg~}\kern 0.1em{\sc ii}~$\lambda\lambda 2796, 2803$}
\def\NVdblt{{\rm N}\kern 0.1em{\sc v}~$\lambda\lambda 1238, 1242$}  
\def\OVIdblt{{\rm O}\kern 0.1em{\sc vi}~$\lambda\lambda 1031, 1037$}
\def\SiIVdblt{{\rm Si~}\kern 0.1em{\sc iv}~$\lambda\lambda1394, 1403$}
\def\AlIIIdblt{{\rm Al~}\kern 0.1em{\sc iii}~$\lambda\lambda1855,1863$}
\def\FeIIdblt{{\rm Fe~}\kern 0.1em{\sc ii}~$\lambda\lambda 2383, 2600$}

\def\AlII{\hbox{{\rm Al~}\kern 0.1em{\sc ii}}}
\def\AlIII{\hbox{{\rm Al~}\kern 0.1em{\sc iii}}}
\def\CaI{\hbox{{\rm Ca}\kern 0.1em{\sc i}}}
\def\CaII{\hbox{{\rm Ca}\kern 0.1em{\sc ii}}}

\def\CrII{\hbox{{\rm Cr~}\kern 0.1em{\sc ii}}}
\def\CII{\hbox{{\rm C~}\kern 0.1em{\sc ii}}}
\def\CIII{\hbox{{\rm C~}\kern 0.1em{\sc iii}}}
\def\CIV{\hbox{{\rm C~}\kern 0.1em{\sc iv}}}
\def\CV{\hbox{{\rm C}\kern 0.1em{\sc v}}}

\def\HI{\hbox{{\rm H~}\kern 0.1em{\sc i}}}
\def\HII{\hbox{{\rm H~}\kern 0.1em{\sc ii}}}

\def\Lya{\hbox{{\rm Ly}\kern 0.1em$\alpha$ }}
\def\Lyb{\hbox{{\rm Ly}\kern 0.1em$\beta$}}
\def\Lyg{\hbox{{\rm Ly}\kern 0.1em$\gamma$}}
\def\Lyfive{\hbox{{\rm Ly}\kern 0.1em$5$}}
\def\Lysix{\hbox{{\rm Ly}\kern 0.1em$6$}}
\def\Lyseven{\hbox{{\rm Ly}\kern 0.1em$7$}}
\def\Lyeight{\hbox{{\rm Ly}\kern 0.1em$8$}}
\def\Lynine{\hbox{{\rm Ly}\kern 0.1em$9$}}
\def\Lyten{\hbox{{\rm Ly}\kern 0.1em$10$}}
\def\HeI{\hbox{{\rm He}\kern 0.1em{\sc i}}}
\def\HeII{\hbox{{\rm He}\kern 0.1em{\sc ii}}}
\def\FeI{\hbox{{\rm Fe~}\kern 0.1em{\sc i}}}
\def\FeII{\hbox{{\rm Fe~}\kern 0.1em{\sc ii}}}
\def\FeIII{\hbox{{\rm Fe~}\kern 0.1em{\sc iii}}}
\def\MnII{\hbox{{\rm Mn}\kern 0.1em{\sc ii}}}
\def\MgI{\hbox{{\rm Mg~}\kern 0.1em{\sc i}}}
\def\MgII{\hbox{{\rm Mg~}\kern 0.1em{\sc ii}}}
\def\MgIII{\hbox{{\rm Mg~}\kern 0.1em{\sc iii}}}
\def\MgIV{\hbox{{\rm Mg~}\kern 0.1em{\sc iv}}}
\def\NaI{\hbox{{\rm Na}\kern 0.1em{\sc i}}}
\def\NV{\hbox{{\rm N}\kern 0.1em{\sc v}}}
\def\NII{\hbox{{\rm N}\kern 0.1em{\sc ii}}}
\def\NIII{\hbox{{\rm N}\kern 0.1em{\sc iii}}}
\def\NiII{\hbox{{\rm Ni~}\kern 0.1em{\sc ii}}}
\def\OVI{\hbox{{\rm O}\kern 0.1em{\sc vi}}}
\def\OI{\hbox{{\rm O}\kern 0.1em{\sc i}}}
\def\OII{\hbox{[{\rm O}\kern 0.1em{\sc ii}]}}
\def\SiII{\hbox{{\rm Si~}\kern 0.1em{\sc ii}}}
\def\SiIII{\hbox{{\rm Si~}\kern 0.1em{\sc iii}}}
\def\SiIV{\hbox{{\rm Si~}\kern 0.1em{\sc iv}}}
\def\SII{\hbox{{\rm S}\kern 0.1em{\sc ii}}}
\def\SIII{\hbox{{\rm S}\kern 0.1em{\sc iii}}}
\def\SIV{\hbox{{\rm S}\kern 0.1em{\sc iv}}}
\def\TiII{\hbox{{\rm Ti}\kern 0.1em{\sc ii}}}
\def\ZnII{\hbox{{\rm Zn~}\kern 0.1em{\sc ii}}}

\def\swift{\emph{Swift \,}}

\def\xray{X-ray}
\def\swift{\emph{Swift}}

\def\simlt{\mathrel{\hbox{\rlap{\hbox{\lower4pt\hbox{$\sim$}}}\hbox{$<$}}}}
\def\simgt{\mathrel{\hbox{\rlap{\hbox{\lower4pt\hbox{$\sim$}}}\hbox{$>$}}}}

\newcommand{\gp}{\mbox{$g^{\prime}$}}
\newcommand{\rp}{\mbox{$r^{\prime}$}}
\newcommand{\ip}{\mbox{$i^{\prime}$}}
\newcommand{\zp}{\mbox{$z^{\prime}$}}

\newcommand{\nh}{\mbox{$N_{\rm H}$}} 

\newcommand{\inex}{\emph{internal-external}}

\begin{document}


\title{Constraining GRB Emission Physics with Extensive Early-Time, Multiband Follow-up}


\author{A.~Cucchiara\altaffilmark{1, 2}, 
	S.~B. Cenko\altaffilmark{1}, 
	J.~S. Bloom\altaffilmark{1}, 
        A. ~Melandri\altaffilmark{3,4}, 
	A.~Morgan\altaffilmark{1}, 
        S.~Kobayashi\altaffilmark{4}, 
         R.~J. Smith\altaffilmark{4}, 
         D.~A. Perley\altaffilmark{1}, 
	W.~Li\altaffilmark{1}, 
	J.~L. Hora\altaffilmark{5},  
	R.~L. da Silva\altaffilmark{2,9},  
	J.~X. Prochaska\altaffilmark{2}, 
	P.~A. Milne\altaffilmark{6}, 
	N.~R. Butler\altaffilmark{1}, 
	B.~Cobb\altaffilmark{7}, 
	G.~Worseck\altaffilmark{2}, 
	C.~G. Mundell\altaffilmark{4}, 
	I.~A. Steele\altaffilmark{4}, 
         A.~V. Filippenko\altaffilmark{1}, 
	M.~Fumagalli\altaffilmark{2},  
	C.~R. Klein\altaffilmark{1}, 
	A.~Stephens\altaffilmark{8}, 
	A.~Bluck\altaffilmark{8},  
	R.~Mason\altaffilmark{8} 
}
\email{acucchia@ucolick.org}

\altaffiltext{1}{Department of Astronomy, University of California, Berkeley, CA 94720-3411}
\altaffiltext{2}{Department of Astronomy and Astrophysics, UCO/Lick Observatory, University of California, 1156 High Street, Santa Cruz, CA 95064, USA}
\altaffiltext{3}{INAF, Osservatorio Astronomicodi Brera, via E. Bianchi 46, I-23807 Merate (LC), Italy}
\altaffiltext{4}{Astrophysics Research Institute, Liverpool John Moores, University, Twelve Quays House, Egerton Wharf , Birkenhead, CH41 1LD}

\altaffiltext{5}{Harvard-Smithsonian Center for Astrophysics, 60 Garden St., Cambridge, MA 02138}

\altaffiltext{6}{University of Arizona, Steward Observatory, 933 N. Cherry Ave., Tucson, AZ 85719}
\altaffiltext{7}{George Washington University}

\altaffiltext{8}{Gemini Observatory, 670 North A'ohoku Place, Hilo, HI 96720}
\altaffiltext{9}{NSF Graduate Research Fellow}

\begin{abstract}


Understanding the origin and diversity of emission processes responsible for Gamma-ray Bursts (GRBs) 
remains a pressing challenge. While prompt and contemporaneous panchromatic observations have 
the potential to test predictions of the internal-external shock model, extensive multiband imaging has 
been conducted for only a few GRBs.  We present rich, early-time, multiband datasets for two \swift\ events, GRB
110205A and GRB 110213A. The former shows optical emission since the early stages of the prompt phase, 
followed by the steep rising in flux up to $\sim 1000$~s after the burst ($t^{-\alpha}$ with $\alpha=-6.13 \pm 0.75$). 
We discuss this feature in the context of the reverse-shock scenario and interpret the 
following single power-law decay as being forward-shock dominated. Polarization measurements, 
obtained with the RINGO2 instrument mounted on the Liverpool Telescope, also provide hints on 
the nature of the emitting ejecta. The latter event, instead, displays a very peculiar optical to near-infrared
lightcurve, with two achromatic peaks. In this case, while the first peak is probably due
to the onset of the afterglow, we interpret the second peak to be produced by newly injected material,
signifying a late-time activity of the central engine.

\end{abstract}



\section{Introduction }
\label{sec:intro}
Despite the unassailable utility of GRBs as probes of the Universe \cite[e.g.][]{tfl09,Totani:2006bl,sdc09,Cucchiara:2011uq}, some basic questions about the nature of the emission mechanisms persist. The \inex\ shock paradigm, whereby the prompt gamma rays arise from self-shocking of an unsteady relativistic wind \citep{Kobayashi:2003ul,Sari:1997ys} and the afterglow arises from shock interaction with the ambient medium \citep{Zhang:2003dq}, has found support \citep[e.g. ][]{Guidorzi:2011yq,Shao:2005vn} and challenges \citep{Zhang:2011kx} from observations. The role of the \emph{reverse}-shock --- that crossing back through the ejecta  --- in the dynamics and observables remains largely unconstrained owing to the lack of good early time multicolor observations when the reverse-shock contribution should be most prominent. In the pre-\swift\ era only a handful of GRBs were detected with sufficient temporal resolution, but an unambiguous case of reverse-shock contribution was not established \citep{Kobayashi:2000lq, Zhang:2003dq,Gruber:2011rr,Perley:2008fk,Gomboc:2008qy,Mundell:2007lr}.
The \swift\ satellite  \citep{Gehrels:2004fj} has permitted unprecedented observations of GRBs at early times \citep{Vestrand:2006pd,Klotz:2008yq}, allowing rapid investigation of emission from hard X-ray to optical frequency regimes. Similarly, fast-response, robotic ground-based telescopes have been optimized in order to quickly follow-up \swift\ events. Thanks to a trigger-rate of 100 events/year, a
large variety of GRBs have been observed, showing several different characteristics. 
For example, afterglows have shown more transitional phases in their lightcurves than pre-\swift\ samples \citep{Evans:2009uq}. In some cases, they show optical and X-ray flares, indicating a refreshing activity from the central engine \citep{Falcone:2007ly}.
All of this new information has made the quest of a ``standard-model'' very challenging, and, after six years of investigation, is still not fully understood.

In this paper we detail high-energy \swift\ observations of GRB 110205A and GRB 110213A,
and associated ground based observations from several facilities typically starting around the end of the prompt emission.
 Multiband observations, spanning several orders of magnitude in frequency and time,  
 in combination with afterglow spectroscopy and host galaxy imaging, 
 represent two extensive datasets in order to further
investigate  the \emph{reverse}-shock emission \citep[see also][]{Gao:2011ul}.
In \S \ref{sec:samp05A} and \S \ref{sec:samp13A} we present our datasets and 
the data analysis procedure; in \S \ref{sec:results} we show our light curves and spectral energy distribution modeling analysis, and the uncertainties involved in the theoretical modeling. 
Finally, \S \ref{sec:discussion} will summarize our finding and discuss some of the future 
prospects of GRB investigation.
Throughout the paper all errors are quoted as a  90\%
 confidence interval, unless otherwise noted. We use a  standard cosmological model 
with $H_0 = 71$ km/s/Mpc, $\Omega_M = 0.27$, and
 $\Omega_{\Lambda} = 0.73$.   


\section{GRB 110205A Data Set}
\label{sec:samp05A}


\subsection{Space-based data}
\label{sec:data05A}
GRB 110205A was discovered by the 
\swift\ satellite on 2011 February 5
at $T_0 =$ 02:02:41 (UT dates are 
used throughout this paper). The BAT instrument \citep{Barthelmy:2005lr}
showed a complex structured emission in the $15 - 350$ keV
energy range \citep{Beardmore:2011fk}.
The GRB  lightcurve shows several overlapping peaks 
 rising around $T_0-120$~s, with the tallest peak at $T_0+210$~s, 
 and minor activity until $T_0+1500$~s.
The duration of the main pulse,
measured in the $15-350$ keV energy range, 
was $T_{90} = 257 \pm25$~s \citep{Markwardt:2011lr}.

At the same time, the {\it Suzaku}-WAM 
all-sky monitor observed the 
emission from this object in the $20-3000$ keV energy range. 
Combined with the BAT data we constructed  
 a  joint spectrum in the energy range from 15 to 3000 keV,
which is well fit by a power-law with exponential cutoff 
model ($dN/dE \sim E^{\Gamma_{\gamma}}\times e^{-(2+\Gamma_{\gamma})E/E_{peak}}$).
The best-fit spectral parameters are 
$\Gamma_{\gamma}=-1.59_{-0.06} ^{+0.07}$ and $E_{peak}=230_{-65}^{+135}$ keV.
The energy fluence in the $15-3000$ keV band calculated for
this model is $2.7_{-0.4}^{+0.7}\times 10^{-5}$ erg cm$^{-2}$
\citep{Sakamoto:2011aa}. Assuming this value and 
a redshift of $z=2.22$ for GRB 110205A (see Sec \ref{sec:spectra}) the 
isotropic-equivalent energy released is $E_{iso} = 4.34_{-0.7}^{+0.4} \times10^{53}$ 
erg in 1 keV to 10 MeV range \citep{PalShin:2011fj}.

\begin{figure*}
\epsscale{1.0}
\includegraphics[scale=0.60,angle= 0]{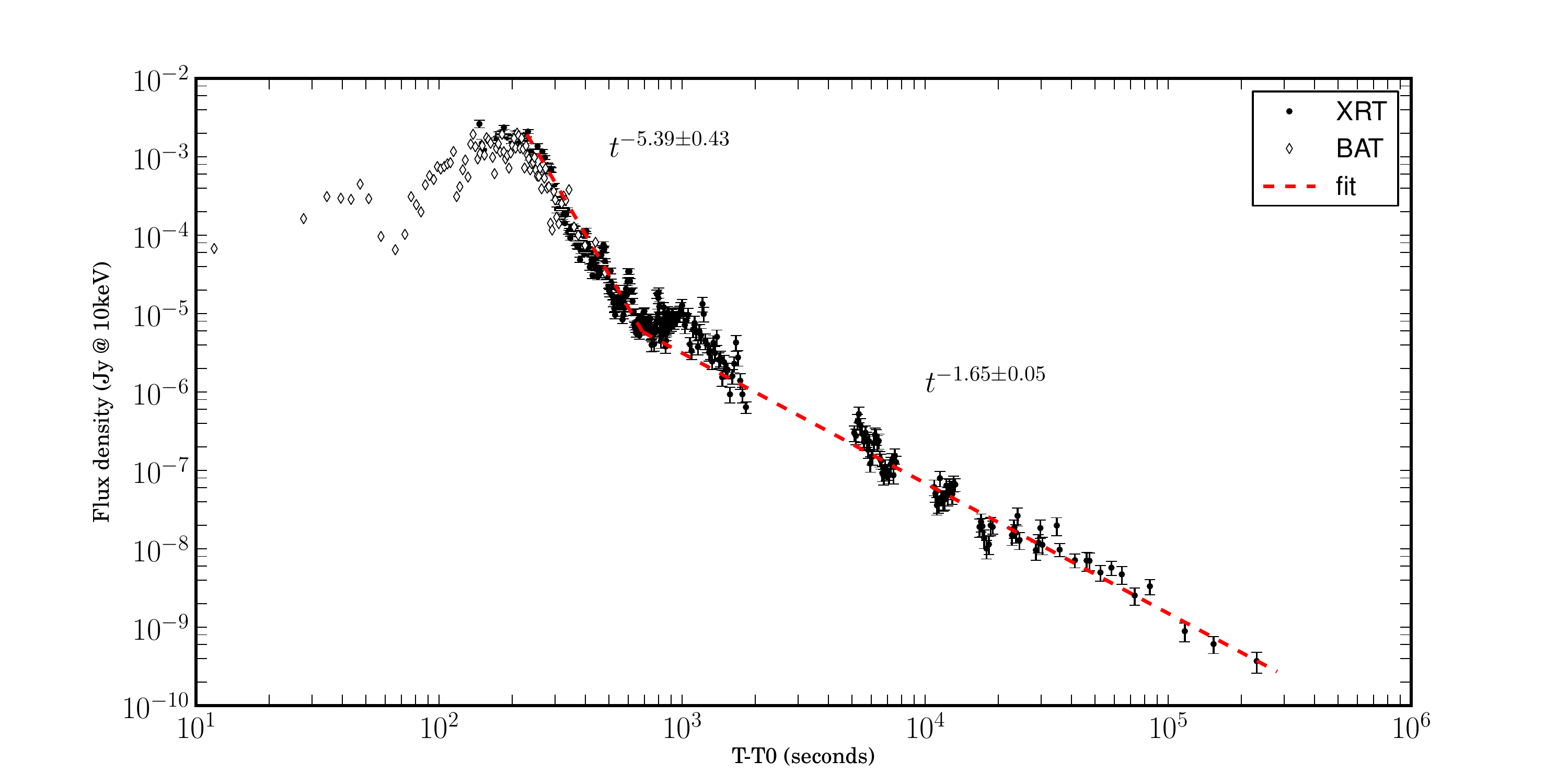}

\caption{GRB110205A BAT (open diamonds) and XRT (filled circles) lightcurve. The BAT emission is rescaled to the XRT energy bands
using a spectral index $\Gamma_{\rm ave}=1.59$ (see \S \ref{sec:1205prompt}). Dashed lines indicate the different power-law segments obtained by fitting the XRT data with single power-laws.}
\label{fig:05Axrtlc}
\end{figure*}

The \xray\ Telescope \citep[XRT;][]{Burrows:2005fk} started 
observing the field 155.4~s after the trigger and showed an uncatalogued fading source 
at \xrtpos\ (J2000), with an uncertainty of $1.5''$ \citep[90\% confidence, ][]{Beardmore:2011lr}.
The XRT and BAT data were obtained from publicly 
available repositories \citep{Evans:2007fj,Evans:2009uq,Butler:2007rt, Butler:2007vn}.
At early times ($T \lesssim 10^{3}$ s; note that henceforth, $T$ is often
used as a shorthand substitute for $T - T_0$),
the ligthcurve displays complex behavior, due probably to the temporal overlap of different pulses similar to 
the pulses identified in the BAT data, as well as flaring activity.
After the prompt phase, the \xray\ afterglow declines until $\sim 600$ s with a steep power law ($\alpha_X=5.39\pm 0.43$). 
Finally, the late-time behavior ($T \gtrsim 10^{3}$ s) can be fitted by a 
single power-law with decay index $\alpha_{X}=1.65\pm 0.05$.
A double power-law with decay indices $\alpha_{X,1}=1.67 \pm 0.05$ and $\alpha_{X,2} =2.01\pm0.34$ and 
a possible break time around  $\approx80$ ks after the GRB explosion provides a slightly better fit, but
not statistically significant. 
Some flares are clearly visible throughout the \xray\ lightcurve, likely indicating a
continuation  of the progenitor activity (see Figure \ref{fig:05Axrtlc}), as
previously seen in other GRBs \citep{Falcone:2007ly}.

Using the early time data (``window-timing'' mode), the XRT averaged spectrum can be fitted
by an absorbed power-law model with photon index $\Gamma_X=1.42 \pm0.02$,
while the late-time spectrum in the XRT energy bands can be fitted with an
absorbed power-law with a photon spectral index of $\Gamma_X =1.99_{-0.07}^{+0.08}$.
In this case, the best fitted absorption column density, at $z=2.22$, is $\nh =3.5_{-1.5}^{+1.6} \times 10^{21}$ cm$^{-2}$, in addition to the Galactic value of $1.6 \times 10^{20}$
cm$^{-2}$ \citep{Kalberla:2005fk}.

Other space-based facilities provided additional high-energy coverage. 
The {\it Konus}-Wind experiment detected GRB 110205A in the $20 -1200$ keV energy range 
with a 4.5-sigma detection of a possible pre-cursor at $T_0 - 1360$ s. 
In addition, a soft tail up to $T_0+1200$ s has been detected, consistent with a 
similar detection by the Swift-BAT instrument. 
The total fluence is $S= 3.66 \pm 0.35 \times 10^{-5}$ erg cm$^{-2}$ in the 
20-1200 keV range \citep{Golenetskii:2011aa,PalShin:2011fj}.

\begin{figure*}
\epsscale{1.0}
\includegraphics[scale=0.80,angle= 0]{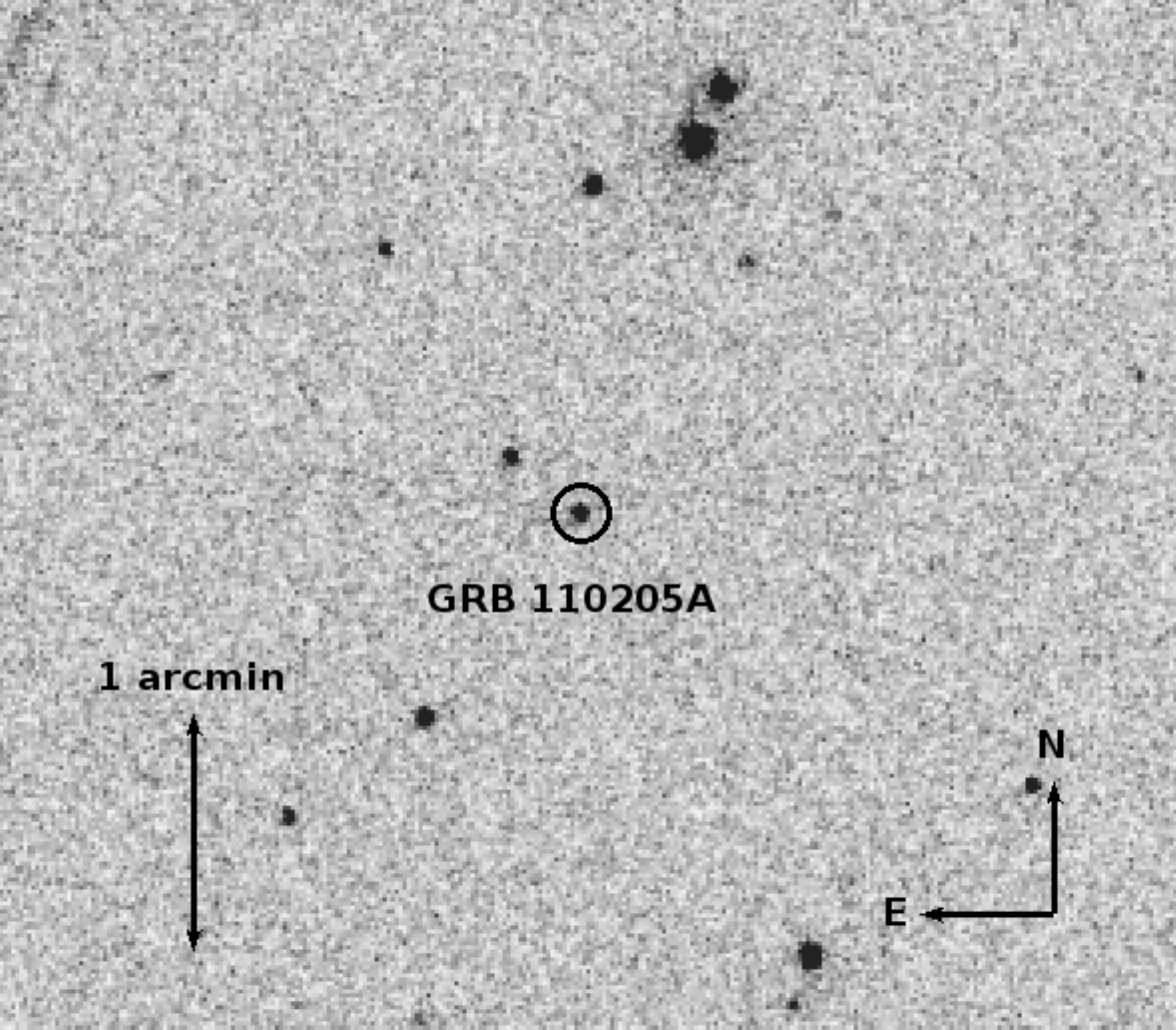}
\caption{UVOT $v$ band image of GRB 110205A obtained $\sim1400$ s after the trigger.}
\label{fig:fin05A}
\end{figure*}

Finally, 164 seconds after the BAT detection, the UVOT instrument \citep{Roming:2005qy}
 onboard \swift\ began observing GRB 110205A, identifying the source in the $white$ band filter
at \uvotRA\ and \uvotDec\ (J2000) with a 0\farcs63 accuracy in both directions \citep[Figure \ref{fig:fin05A},][]{Beardmore:2011lr}.
UVOT early time observations have been acquired in ``image-event" mode, 
allowing a very detailed  time-resolved analysis of the count rate
variation in the UVOT detector.
Using the {\tt HEASOFT} tools {\tt UVOTEVTLC} and 
{\tt UVOTMAGHIST} we estimated the
total flux inside a 5\arcsec\ region centered at the object position, while 
using an annular background region as suggested by \cite{Poole:2008yq} and \cite{Breeveld:2011kx}. Data in the $white$ filter were binned
with a bin size of 8~s in order to reach a minimum signal-to-noise ratio (S/N)
of 3 per bin, in particular during the prompt emission.

The afterglow was also detected in the $u,b,v$ and $w1$ filters up to $T=8\times 10^4$~s.
The lack of detection in the other two, bluer, filters available on UVOT is consistent
with the observed redshift.

\subsection{Ground-based follow-up}
\label{sec:ground}

Soon after the \swift\ trigger, several robotic facilities pointed at the GRB location
providing a series of photometric observations from optical to near-infrared (NIR) and Radio bands.
Our group first identified an infrared counterpart using the 
 Peters Automated Infrared Imaging Telescope \cite[PAIRITEL, ][]{Bloom:2006ij}
which consists of the 1.3 m Peters Telescope at Mt.
Hopkins, AZ, formerly used for the Two Micron All Sky Survey \cite[2MASS;][]{Skrutskie:2006ys},
refurbished with the southern 2MASS camera.
Observations began at
05:14:03, $\sim3.2$ hr after the trigger, and continued 
until the source reached its hour-angle limit. Further
observations were obtained on the following day \citep{Morgan:2011lr,Morgan:2011fk}.  
The raw data files were processed using a standard IR reduction technique 
via PAIRITEL
Pipeline III and resampled using SWarp\footnote{See http://www.astromatic.net/software/swarp.} 
\citep{Bertin:2002fr} to create 1.0\arcsec pixel$^{-1}$ images for final
photometry.

The standard observing mode is to take three 7.8 s exposures
in immediate succession at each dither position. These are then
median-combined into 23.4-s ``triplestacks,'' which were subsequently
binned iteratively until a large enough S/N was achieved at the source
position for accurate photometry. Aperture photometry was performed
using custom Python software, utilizing Source Extractor
 \citep[SExtractor;][]{Bertin:1996zr}
as a back end. Eight calibration stars present
in all images were chosen based on brightness, proximity of nearby
contaminating sources, and location relative to bad pixels. The
optimal aperture of 4\arcsec\ diameter was determined by minimizing the
absolute error relative to 2MASS magnitudes of our calibration stars.

Calibration was performed by redetermining the zeropoint for each
image individually by comparison to 2MASS magnitudes using these eight
stars. The resulting statistical uncertainty in the zero-point is
negligible relative to other sources of error. Additional, systematic
sources of error are addressed in detail by \cite{Perley:2010mz}; we
use a similar procedure here to determine the total uncertainty of
each point.

We also observed the field of GRB\,110205A with the automated Palomar 
60\,inch telescope (P60; \citealt{cfm+06}) 
approximately 96\,minutes after the \textit{Swift}-BAT trigger
time.  After executing a series of pre-programmed $g^{\prime}$, 
$r^{\prime}$, $i^{\prime}$, and $z^{\prime}$ observations, we manually
inserted deeper observations until the afterglow faded below our 
sensitivity threshold.

Basic processing (bias subtraction, flat-fielding, etc.) was performed by 
our custom IRAF\footnote{IRAF is distributed by the National Optical 
Astronomy Observatory, which is operated by the Association for Research 
in Astronomy, Inc., under cooperative agreement with the National Science 
Foundation.} pipeline.  Later images were stacked to increase sensitivity
with the SWarp software.  
Instrumental magnitudes were extracted using point-spread function 
(PSF) fitting routines from the IRAF \texttt{DAOPHOT} package,
and were photometrically calibrated with respect to bright, nearby 
reference stars from the Sloan Digital Sky Survey (SDSS; \citealt{aaa+09e}).

\begin{figure*}
\epsscale{1.0}
\includegraphics[scale=0.50,angle=0]{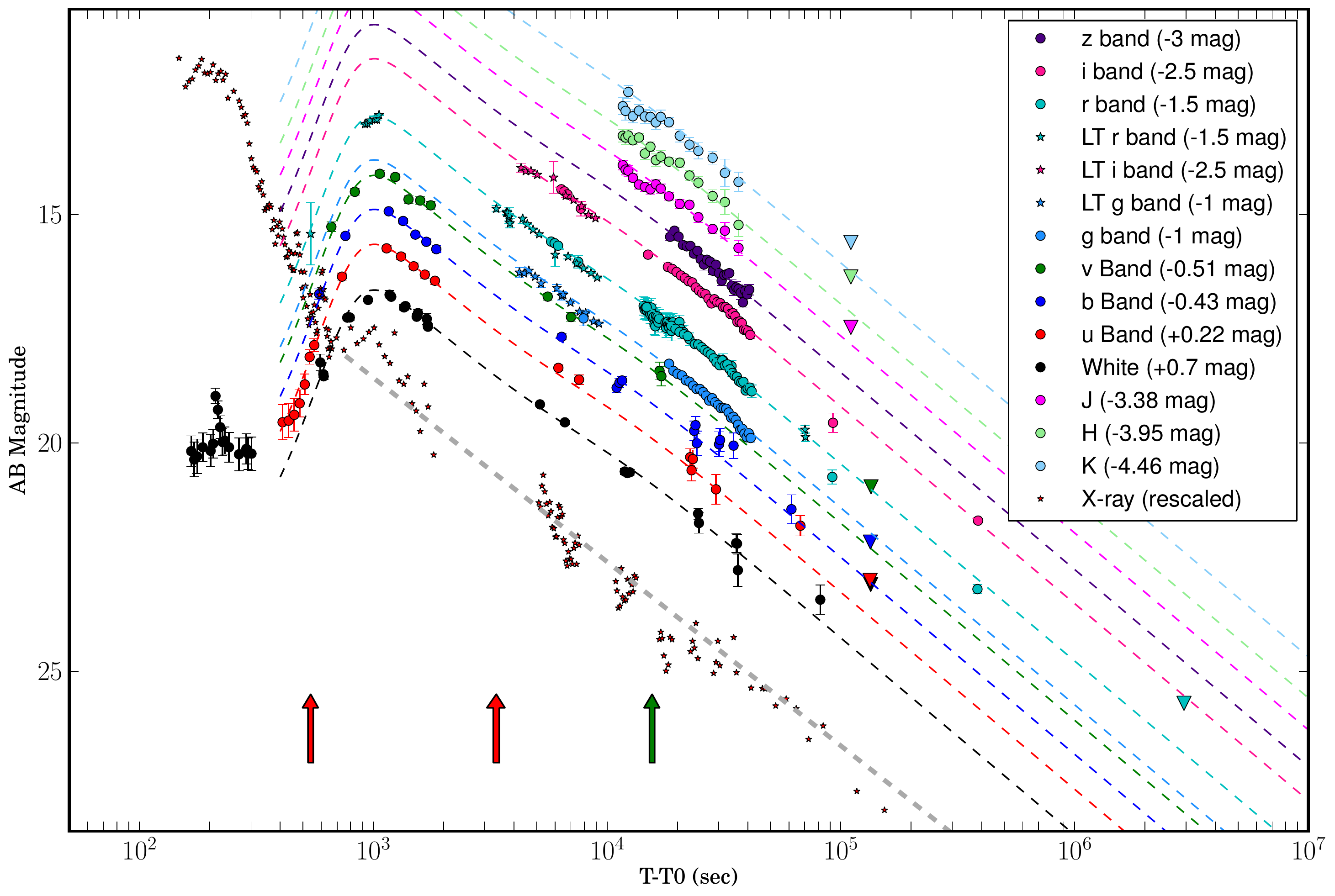}

\caption{GRB110205A lightcurve. We present all the available public data in conjunction with the datasets present in this work. The dashed lines represent a resulting multiband fit (see \S \ref{sec:samp05A} for more details). The XRT light curve is arbitrarily rescaled for comparison and fitted with a simple power-law function. The achromatic steep rise in the optical bands is interpreted as the signature of the reverse shock. After the peak at $T\approx 985$~s the forward shock is the main source of radiation, confirmed by the net polarization measurement obtained at the time of the RINGO2 and the CAFOS instruments observations, indicated with red and green arrows, respectively.}
\label{fig:fig1}
\end{figure*}

Multiband observations of the optical afterglow of GRB 110205A were also, acquired with the 
robotic 2-m Liverpool Telescope (LT) starting about 920~s after the trigger time. After taking a 
sequence of short exposures with the $\rp$ filter ($6 \times 10$~s), the LT continued to monitor 
the field cycling through $\gp,\rp$, and $\ip$ filters, using increasing exposures times, until finishing 
about 155~min after the burst event. 
Two deeper, 300 s long, observations in the $\rp$ filter were acquired manually at the 
end of the night, when the OT was still clearly detected (see Figure~\ref{fig:fig1}).
These observations were all photometrically calibrated using the same standard stars 
adopted for the Palomar data.

After $\sim 4.5 $ days, when the afterglow faded beyond the detectability of these
facilities, we activated our Target of Opportunity (ToO) program at the Gemini-North
telescope (P.I. Cobb), in order to monitor the late time behavior of the 
optical transient and/or estimate the possible host galaxy contribution. 
We performed a series of 10x3 minutes exposure in
\rp\ and \ip\ bands using the GMOS camera \citep{Hook:2004fj}. 
We analyzed and coadded these datasets using the
dedicated {\tt GEMINI} tool included in the IRAF environment.
Calibration was performed using calibration stars present in the
P60 data and magnitude were estimated using Source Extractor.
The afterglow was detected and its brightness in the two bands is consistent
with the extrapolation from early-time data, following a single power-law 
behavior.

In order to establish a possible host galaxy contribution, we repeated our 
\rp\ band observation on 11 March 2011. We collected a total exposure time 
of 40 minutes. No object is detected at a 3-sigma upper limit of $\rp > 27.21$ mag,
corresponding to a flux density limit of $4.88 \times 10^{-2}\, \mu$Jy, implying no significant 
host contamination in our earlier observations.
A summary of all our observations can be found in
Tables \ref{tab:tab1}--\ref{tab:tab11} and shown in Figure~\ref{fig:fig1} with all our optical 
and infrared data.

The overall lightcurve in the optical and near-infrared bands after the first 300~s
can be fitted by a model with two double power-law components 
 known as ``Beuermann functions''  \citep{Beuermann:1999fk}: 

\begin{center}

\begin{displaymath}
F_{i}(t) = \sum_{n=1,2} \, F_{i,n} \left[ \left( \frac{t}{t_{b_{n}}}\right)^{\alpha_{n}} + \left( \frac{t}{t_{b_{n}}}\right)^{\beta_{n}} \right] ^{-1}. 
\end{displaymath}

\end{center}

The normalization factors ($F_{i,n}$) are different for each of the $i$ datasets and are free parameters in our fitting procedure, as well as the power-low indexes ($\alpha_{n}$ and $\beta_{n}$) and the
break times ($t_{b_{n}}$), which are considered to be the same for all the bands.
In this formalism, the peaks of the individual function can be estimated as:

\begin{center}

\begin{displaymath}
t_{peak,n} = t_{b_{n}}\times (-\alpha_n / \beta_n)^{1/(\beta_n - \alpha_n)}.
\end{displaymath}

\end{center}
\noindent
Both components are needed
in order to account for the steep rise at early time and the exponential decay
after the first peak at $t_{\rm peak} = 985 \pm 48$~s. We fit our datasets simultaneously and the results
are listed in Table \ref{tab:fit05A} as well as presented in Figure \ref{fig:fig1}, before correcting
for Galactic extinction, $E(B-V)=0.02$.

In addition, radio observations were performed with several facilities
providing upper limits in the millimeter and sub-millimeter regimes \citep{Zauderer:2011fr,Petitpas:2011zr,van-der-Horst:2011kx}.
The Expanded Very Large Array facility detected  the radio afterglow 1.2 days after the burst,
providing a flux density of $182 \pm 12$ $\mu$Jy at a frequency of 22 GHz \citep{Zauderer:2011ys}.


%
%
\subsection{Optical Polarization}
\label{sec:05Apol}

Polarization observations were obtained with the RINGO2    
polarimeter on the Liverpool Telescope.
The procedure consists in observing the target object (GRB) and other seven stars
in the field. Also, known unpolarized sources are observed the same night, in order to 
minimize the intrinsic polarization introduced by the polarimeter itself and to 
be able to detect any 
residual polarization in the GRB emission.
A more detailed description of this procedure can be found in \citet{Guidorzi:2006fk} and
\citet{Steele:2010qy}.
The first RINGO2 image was obtained starting at 02:06:43, 243~s
after the BAT trigger time, during the brightening phase of the afterglow.
The data are consistent with the OT being
unpolarized, but unfortunately, due to significant cloud cover
we were able only to determine an upper limit  of $<16\%$ polarized (3-sigma confidence level).

A second observation, centered at
02:58:07.1, 56~min after the trigger, was performed under
significantly better conditions. 
We measure the $V$-plus-$R$-band linear polarization of the OT as 3.6\%, with
a 2-sigma confidence interval of 0 - 6.2\%. 
By randomizing the time sequence of values in the observed trace, we simulate an unpolarized data
stream which has exactly the same photometric characteristics as our observations.
From many such random realizations, we reject the unpolarized hypothesis
at a confidence of 92\%.

These values are consistent with the net optical linear polarization estimated in $R$ band 
few hours after the burst at the level of $P \sim 1.4 \%$, performed by the Calar Alto Observatory 
equipped with the CAFOS instrument \citep{Gorosabel:2011uq}. Both observations 
are indicated as red and green arrows in Figure \ref{fig:fig1}.

%
%

\subsection{Absorption Spectroscopy}
\label{sec:spectra}
We began observing the optical afterglow of GRB\,110205A with the 
FAST spectrograph \citep{fcc+98} mounted on the 1.5\,m Tillinghast reflector
at Mt.~Hopkins Observatory at 5:11 on 2011 February 5.  We obtained 
two 1800\,s spectra with the 300 line mm$^{-1}$ grating, covering the wavelength 
range 3500--7500~\AA\ \citep{Cenko:2011vn}.  
The data were reduced using standard IRAF
routines, including optimal extraction and wavelength calibration relative
to a series of HeNeAr calibration lamps.  Flux calibration was performed
relative to the standard star Feige\,34.  The resulting coadded, normalized 
spectrum is shown in Figure~\ref{fig:spectrum05A}.

Another spectrum \citep{da-Silva:2011rt} was taken with the Kast spectrograph \citep{Miller:1993ms} on the 3~m Shane reflector at Lick Observatory.
The reduction procedure was the same as the previous one and the resolution
of this spectrum is $\sim4$ \AA\ in the blue side (around 4500 \AA) and $\sim10$ 
\AA\ in the red side (around 6500 \AA).
This spectrum shows a prominent damped Ly$\alpha$ (DLA) absorption
 system as observed in other GRBs.
This feature, in combination with other metal lines (e.g. \FeII, \SiII\ and \SiII*, \CIV) place 
the GRB at $z = 2.21442 \pm 0.00044$ (Figure \ref{fig:DLA05A}).
Equivalent widths are estimated  and listed in Table \ref{tab:05Alines}.
The neutral hydrogen column density, estimated by fitting the DLA with a Voigt profile, is
log$(\NH)=21.45 \pm 0.20$,
consistent with the one derived by the X-ray analysis, likely implying 
 minimum photoionization of the hydrogen in the
circumburst material caused by the burst radiation field \citep{Campana:2010fk,Watson:2007ly}.

\section{GRB 110213A Data Set}
\label{sec:samp13A}

\label{sec:samp13A}
\subsection{Space-based Data}
GRB 110213A was discovered by  \swift\ on February 13 at 05:17:29. 
The BAT lightcurve has a typical single-pulse  shape, with $T_{90} = 48 \pm 16$ s,
estimated in the 15--350 keV energy range. 
The time-averaged spectrum from $T_0 -31.2$ to $T_0 +32.8$ s is best fit by a simple 
power-law model with a power-law index of $\Gamma_{\gamma}=1.83 \pm 0.12$.
The fluence in the 15--150 keV band is $5.9\pm 0.4 \times 10^{-6}$ erg cm$^{-2}$ 
\citep{DElia:2011ly,Barthelmy:2011ve,Stratta:2011ul}.
The {\it Konus}-Wind experiment also observed this event reporting similar results \citep{Golenetskii:2011qy}.
Finally, the Gamma-ray Burst Monitor (GBM), onboard the \emph{Fermi} satellite
detected the prompt emission of this event \citep{Foley:2011uq}.
In the energy range 50--300 keV, the spectrum is well fit by a power-law function
with exponential cut-off. The power-law index is $\Gamma_{\gamma} = -1.44 \pm 0.05$ 
and $E_{\rm peak}=98.4_{-6.9}^{+8.6} $ keV. Using the observed fluence value of
$1.03 \pm 0.03 \times 10^{-5} $ erg cm$^{-2}$ and 
$z=1.46$ (see \S \ref{sec:13Aspec}) we derive an isotropic-equivalent energy of 
$E_{\rm iso}=7.2^{+0.1}_{-0.08} \times 10^{52}$ erg.


The \swift\ spacecraft slewed immediately, allowing the
XRT to be on target in $\sim81$~s and
to continue observing up to $\sim50$~ks after the GRB discovery.
The XRT enhanced position of the afterglow is \xrtposratwo, \xrtposdectwo, with 
an uncertainty of 1.5 arcsec in both directions \citep{Osborne:2011fk}.
The XRT light curve can be modeled by different power-law components ($t^{-\alpha}$):
initially, at $ T < 200$ s, the afterglow steeply decays proportional to $\alpha=4.96\pm0.21$.
Subsequently, it flattens and then slowly  rises as $\alpha=-0.44\pm0.10$ until $T \approx1500$~s 
when it starts fading again with $\alpha=1.08\pm0.04$. A final steepening occurs
at $T  \approx 10^4$ s, after which the \xray\ decay as $\alpha=1.98\pm0.04$. 
There are hints of a possible jet break around $\sim 1$ day postburst, based on the last 
observation, which, if valid, would place useful constraints on geometry of the burst 
emitting region (Figure \ref{fig:13Axrtlc}).

UVOT started observing  $\sim100$ seconds after the BAT trigger.
The afterglow was detected in  the $white, u,b,v$ and $uvw1$ filters at
\uvotRAtwo, \uvotDectwo, with an uncertainty of 0.61 arcsec in both directions (Figure \ref{fig:fin13A}).
In contrast to the early \xray\ data, no steep decay is detected at $T\lesssim300$ s, and instead 
a rising behavior is present in the $white$ band observations 
with a power-law index $\alpha=-2.08 \pm0.23$. 
Similarly to GRB 110205A, two Beuermann functions fit our datasets simultaneously very well. 
A peak is detected at $T=315 \pm 85$ s, after which the optical/UV emission fades as 
$\alpha=1.10 \pm 0.24$ until a minimum flux point at around 2000 s, when the emission 
steeply brightens again with $\alpha=-2.02 \pm 0.34$. After reaching a second peak 
at $T\approx4900$ seconds, the emission decays as $\alpha=1.80 \pm0.15$. 
Unfortunately, due to the afterglow faintness, the UVOT data are sparse, but still
cover up to $50$ ks after the burst.

\subsection{Ground based follow-up}
\label{sec:ground13A}
The Katzman Automatic Imaging Telescope \cite[KAIT;][]{Filippenko:2001qy} responded to the trigger within $\sim 70$~s 
after the trigger, observing the  afterglow in the unfiltered band followed by 
$V$ and $I$. The data were analyzed similarly to the P60 data 
(\S \ref{sec:ground}) and calibrated using USNO cataloged field stars.
The afterglow was detected until $T \approx 3000$ s, during the second rise 
seen in the UVOT data.

The P60 began observing the optical afterglow of GRB 110213A
about 162~s after the trigger, and continued sampling the light curve 
beyond $10^5$~s in the 4 optical filters available, complementing the 
KAIT observations and confirming the two-peak 
behavior of the low-energy afterglow emission.


Finally, in order to constrain the late-time ($T > 10^4$~s) behavior we triggered our ToO
program at the Gemini-North telescope (P.I. Cobb), using the GMOS 
camera \citep{Hook:2004fj} in imaging mode, performing \rp\ and \ip\ observations six 
days after the trigger.
The object brightness in these bands indicates a possible jet break, in agreement 
with the \xray\ analysis, around $T \approx 1$ days after the burst.
A complete summary of our observations can be found in Tables \ref{tab:tab12}--\ref{tab:tab20}, and 
the full light curve is shown in Figure \ref{fig:lc13A} before correcting for Galactic extinction, $E(B-V)=0.32$.

\subsection{Absorption Spectroscopy}
\label{sec:13Aspec}
We also determined the redshift of this GRB with the 
Boller and Chivens Spectrograph mounted on the Steward 2.3-m Bok telescope on Kitt Peak, AZ.
Based on several metal lines, including \FeII, \FeII*, \NiII\ and \AlII, we
found $z=1.4607\pm 0.0001$  (see Figure~\ref{fig:spec13A}).
Equivalent widths for some of these features are listed in Table \ref{tab:0213Alines}.


\section{Results}
\label{sec:results}
GRB 110205A and GRB 110213A present well-sampled light curves 
from high energy to optical and NIR bands, covering a large timeline.
We now frame the observed behavior of the afterglow emission within the  \inex\ shock scenario,
emphasizing the analogies and differences as well as the limitations of the current
follow-up efforts. The data allow constraining statements about
the emission mechanisms and the interaction of the afterglow with the surrounding 
material.
Throughout the following we will use the usual notation where $F(\nu,t) \propto \nu^{-\beta} f(t)$,
where $f(t)$ is a Beuermann function or a simple power-law model and $\beta$ the spectral index.

\subsection{GRB 110205A}
GRB 110205A represents an important laboratory in which to test the standard 
\inex\ shock model. It presents several characteristics of
the ``typical'' GRB prompt emission as well as of the afterglow component. 
We can divide the detected emission into 3 main parts: 1) the prompt emission
($T \lesssim 400$ s), which has been observed by all three 
instruments onboard the \swift\ satellite; 2) the optical peak region ($400\lesssim T
\lesssim 10^3$ s after the burst); 3) the late time phase ($T \gtrsim 10^3$ s).

\subsubsection{Prompt phase}
\label{sec:1205prompt}
The prompt emission of GRB 110205A, as detected by the BAT instrument, is composed 
of a series of peaks as seen in several other long GRBs.
The observed fluence, $f= 2.7_{-0.4}^{+0.7}\times 10^{-5}$ erg cm$^{-2}$, 
is on the higher end of  the \swift -GRBs fluence distribution 
at similar redshift as estimated recently by \cite{Meszaros:2011gf}
and the isotropic energy emitted, $E_{iso}$, places GRB 110205A
well within the 3-sigma confidence level of the $E_p - E_{iso}$ correlation
for long GRBs \citep{Amati:2008fy}.

\begin{figure*}
\epsscale{1.2}
\includegraphics[scale=0.50,angle=0]{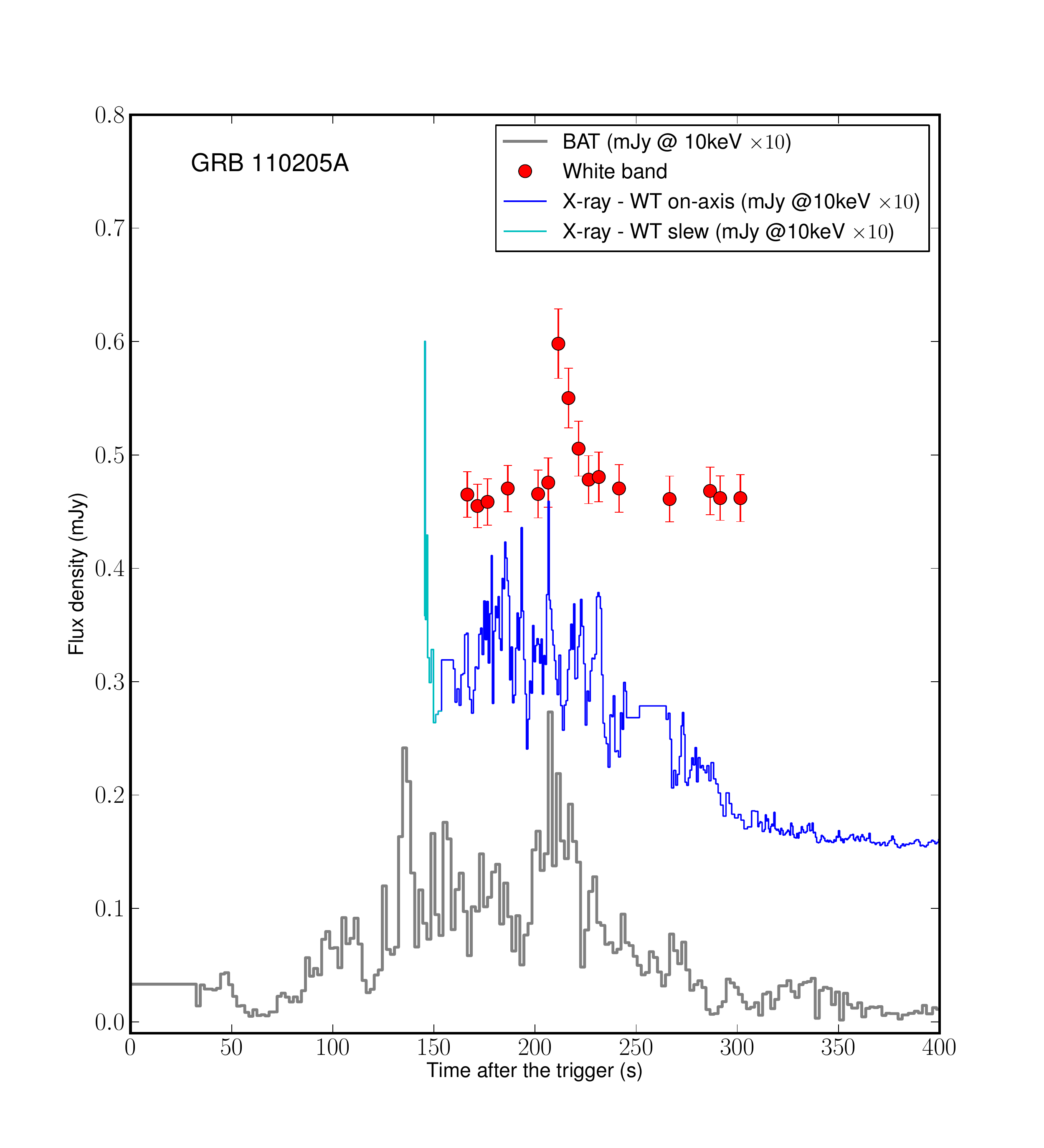}

\caption{Early time light curve comparison. The BAT flux density (grey histogram, 1-sec bin) has been extrapolated in the XRT bandpass (0.3--10 keV). 
Red points are UVOT $white$ band detections (5~s bin), while the cyan and blue points are XRT window-timing mode detections (0.5~s bin) during and after the slewing procedure respectively. There is a hint of correlation between the optical and the high-energy bands, probably indicative of a similar emission mechanism during the prompt phase.The three curves have been shifted along the ordinate to facilitate this comparison.}
\label{fig:fig2}
\end{figure*}

Furthermore, similarly to other cases \citep[e.g. GRB 080319B,][]{Racusin:2008lr}, the XRT and UVOT 
instruments were able to 
begin observing before the end of the prompt phase.
The multiple peak nature of the prompt emission is consistent with each peak being produced by 
internal shocks due to the collision of different shock fronts.
In Figure~\ref{fig:fig2} we overplot the UVOT 
\emph{white} band lightcurve in comparison with the XRT and the BAT
signal during the first 400~s after the burst trigger. A bright and very sharp 
 peak at $\sim 220$s ($\Delta t/t \lesssim1$) 
is detected also in the UVOT band, and it is 
consistent with an internal dissipation process \citep{Ioka:2005rt}.

Over the entire prompt phase, the \xray\ and gamma-ray emissions have a similar average 
photon index ($\Gamma_{\rm ave} \approx 1.59$) derived from the BAT and {\it Suzaku} data.
Rescaling the hard \xray\ emission into the XRT energy range, the lightcurves align with each other as 
can be seen clearly in Figure \ref{fig:05Axrtlc}.

\begin{figure*}
\epsscale{1.0}
\includegraphics[scale=0.60,angle= 0]{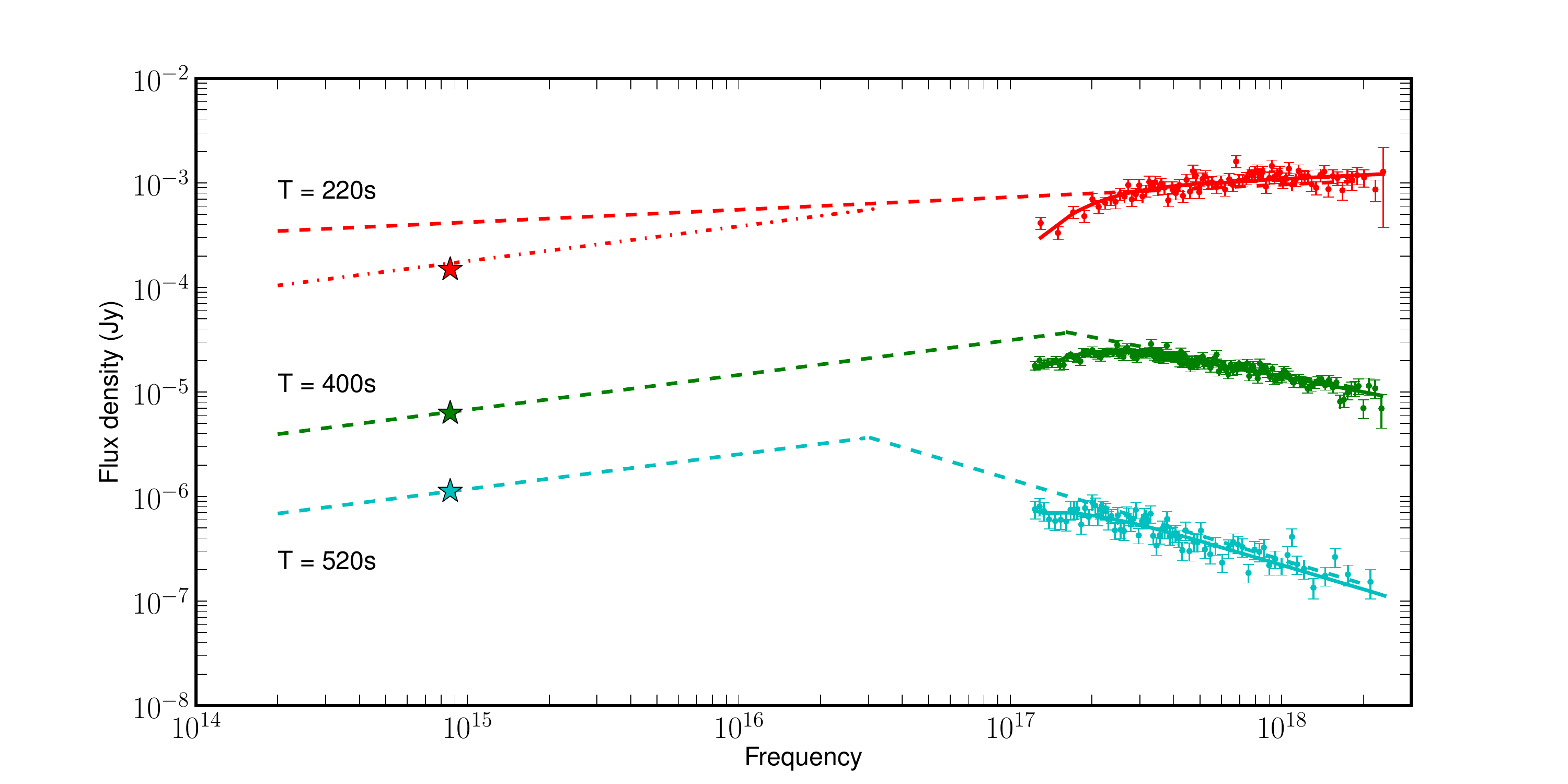}

\caption{Spectral energy distribution constructed using the XRT and the UVOT $white$-band data at the time
of the optical peak ($T \approx 220$~s), at $T \approx 400$~s, and at $T \approx 520$~s, when the high-latitude effect likely dominates the 
observed emission. Fitting the SED at the peak with a broken power law, as shown by the dashed-dotted line (imposing a spectral index $\beta=1/3$ at $\nu_{\rm opt}<\nu<\nu_{\rm X}$), implies
a spectral break which is inconsistent with the spectral evolution at later times. The most likely scenario is that the $white$-band
detection is affected by attenuation due to the DLA and the broad-band transmission curve of the UVOT filter.}\label{fig:sedfig}
\end{figure*}

\begin{figure*}
\epsscale{1.0}
\includegraphics[scale=0.60,angle=0]{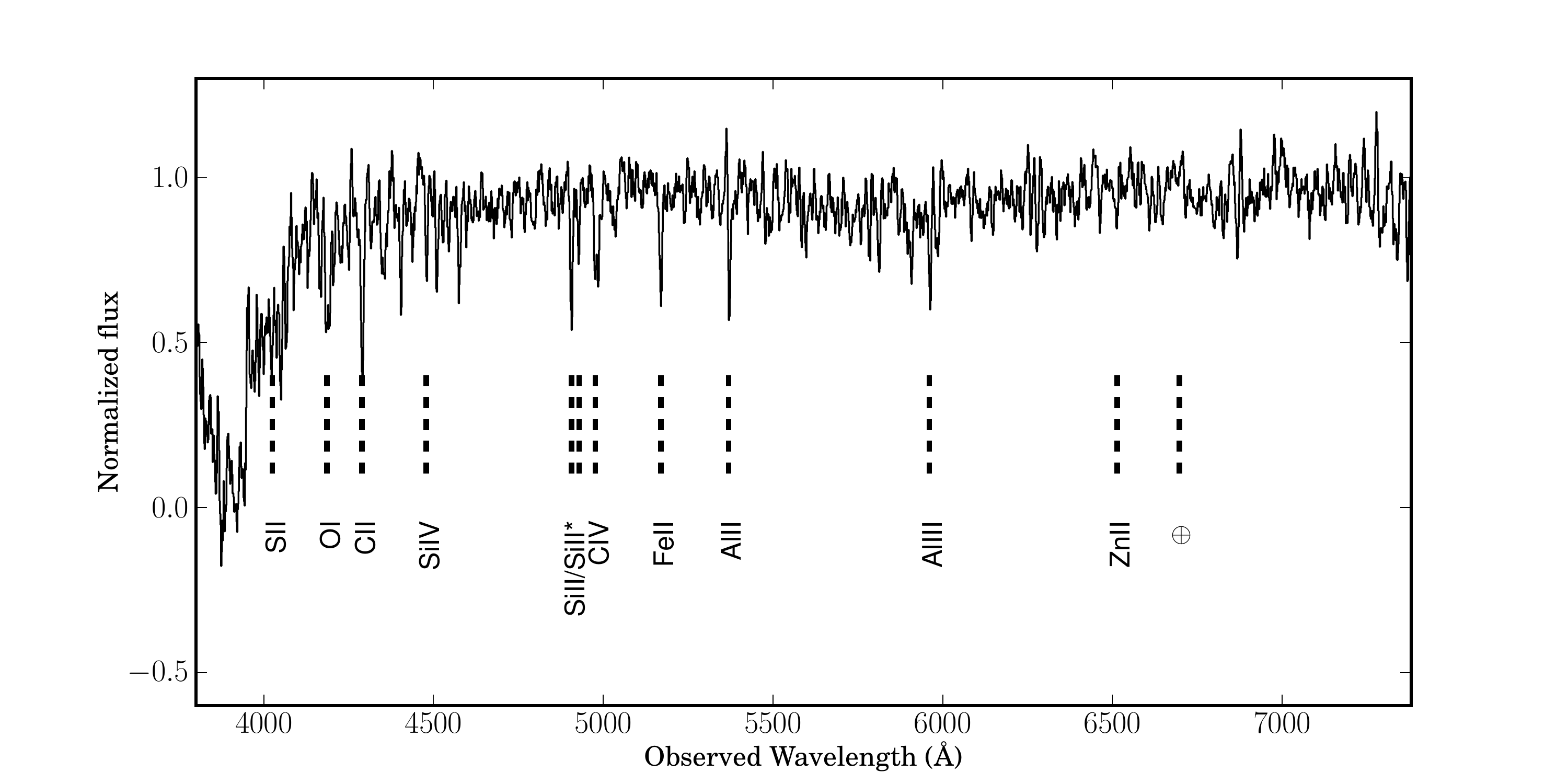}
\caption{Normalized spectrum of the afterglow of GRB 110205A obtained with the FAST spectrograph 
$\sim3.5$h 
after the burst (see Sec. \ref{sec:spectra} for details). The main absorption features are labeled as well the main atmospheric absorption bands.} 
\label{fig:spectrum05A}
\end{figure*}

\begin{figure*}
\epsscale{1.0}
\includegraphics[scale=0.50,angle=0]{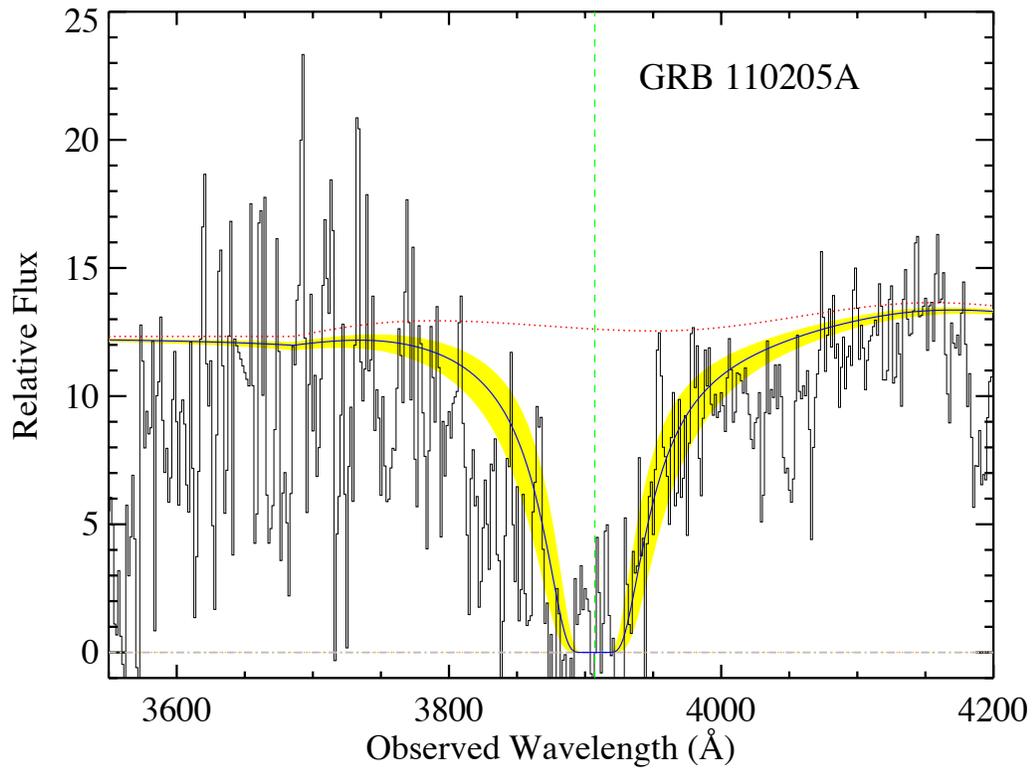}

\caption{Portion of the spectrum obtained with the Kast spectrograph at Lick Observatory. A DLA absorption system is clearly detected, placing 
GRB 110205A at $z=2.22$. The blue line represents the DLA fit, while the shaded area is the 1$\sigma$ confidence interval. 
The dotted red line is the continuum fit and the green dashed line identifies the location of the 1216 \AA\ Ly-$\alpha$ 
feature. A Voigt profile fit gives a neutral hydrogen column density of
log$(\nh/{\rm cm}^{-2}) = 21.45 \pm 0.23$.}
\label{fig:DLA05A}
\end{figure*}

Furthermore, at the most prominent peak in the BAT data, at $T\sim 220$ seconds after burst, the \xray\ spectral index  is  $\beta \sim -0.12 \pm 0.04$.
In Figure \ref{fig:sedfig} we present the spectral energy distribution (SED)
constructed using the \xray\ data and the $white$ filter observation at $T= 220$~s. 
A single power-law extrapolation from the high-energy band ovepredicts the 
$white$-band observation of a factor of $\sim 2$, indicating that the \xray\ and the optical emission
may likely belong to the same segment of the SED and are produced by the same electron population
\citep[see also][for similar studies]{Rossi:2011uq,Shen:2009yq,Vestrand:2006pd}.
The discrepancy in the observed low-energy flux is not surprising, since the broad band filter extends blueward at 2000 \AA\ and redward up to 6000 \AA.
Part of the flux, then, is suppressed by the presence of the Lyman-$\alpha$ break 
which is redshifted at 3912 \AA\  ($z=2.214$).

A break between the \xray\ and the optical bands (imposing $\nu^{\beta}$ with $\beta=1/3$; dash-dotted line 
in the figure) would fit the data better, but would be inconsistent with the SED at later time.
In fact, as comparison, we present a multiband SED constructed at $T\approx400$~s and $T \approx 520$~s
after the burst. 
In these cases a break will be required in order to account for the change in the \xray\ spectral index
as seen in other similar cases \citep{Rossi:2011uq}, probably due to one of the characteristic synchrotron
frequencies.
The following fast decline observed in the \xray\  regime is probably due to the tail of the prompt emission and it is governed by the high-latitude effect \citep[hereafter HLE,][]{Kumar:2000kx}, 
for which emission from different viewing angles reach the observer
with different delays due to light propagation effect \citep[see][and references within for a complete taxonomy of the X-ray lightcurves]{Racusin:2009nx}.
At the same time ($350 \lesssim T\lesssim 600$), the \xray\ spectrum undergoes a hard-to-soft 
evolution which has been characteristic of a large number of GRBs. In the HLE, the temporal and spectral indices are correlated such that $\alpha= 2+\beta$ \citep{Kumar:2000kx}. 
In the case of GRB 110205A, this is not satisfied because the derived values are $\alpha_{\rm X} = 5.39$ and $\beta_{\rm X} \approx 0.5$,
but, as shown in \cite{Zhang:2006lr}, shifting the time zero point ($t_0$) of the afterglow can reconcile
the the observation with the theory without ruling-out the curvature-effect 
interpretation. Assuming that the afterglow starts at $t_0\approx 200$~s a new fit of the \xray\
light curve gives a temporal index of $\alpha_{\rm X} = 2.70 \pm 0.10$, in agreement with the expected
value from  the curvature effect ($\alpha_{\rm theo}=2.5$).

\subsubsection{Optical peak time}
After 600 seconds, the \xray\ afterglow declines with a temporal
index  $\alpha_{\rm X}=1.65$, likely the signature of a \emph{forward}-shock dominated emission taking place. 
Flaring activity is also detected at this time, which is not unusual for the \xray\ 
emission \citep{Margutti:2011rr,Chincarini:2010dq,Gao:2009bh,Marshall:2011lq}.


At the same time, however, the optical emission undergoes a steep increase in flux: assuming that this 
rising begun at the time of the trigger ($t_0=t_{\rm trigger}$) the optical flux increases as
 $t^{-\alpha_{\rm opt}}$, with $\alpha_{\rm opt}=-6.13\pm 0.75$ ($\chi^2/{\rm d.o.f.}=1.29$) 
until $t_{peak}\approx 985$ s. 
As discussed in other cases \citep{Liang:2006lr,Quimby:2006fk,Lazzati:2006qy,Kobayashi:2007lr,Rossi:2011uq} the power-law index is very sensitive to the choice of $t_0$.
In particular, the rise slope is estimated by $d\, {\rm ln}\, F_{\nu}(t) / d\, {\rm ln}\, (t-t'_0)$, where $t'_0=t_{\rm trigger}+t'$, instead of
the usual $d\, {\rm ln}\, F_{\nu}(t) / d\, {\rm ln}\, (t-t_{\rm trigger})$.
Assuming that the optical emission started \emph{after} the actual trigger time ($t'>0$), for example at the
time of the optical brightest point in the $white$ band (see previous section), the fit slightly improves ($\chi^2/{\rm d.o.f.} = 1.19$) 
and a shallower rising index of $\alpha_{\rm opt} = -3.96\pm 0.86$ is obtained. This value is consistent with a reverse-shock
theoretical prediction \citep[e.g., $\alpha_{\rm theo} = -5$; ][]{Zhang:2003dq}. 
Instead, assuming an \emph{earlier} emission ($t'<0$) does not produce a better fit and, as result, implies an even steeper power-law index.
In this last scenario, considering also the absence of a clear precursor in the gamma-ray band, it would be difficult to explain such emission in the optical before the hard \xray.


Overall, this peculiar behavior, similar in all the observed bands from UV to Optical, can originate from a \emph{forward}-shock or, as suggested in a few other cases, from the \emph{reverse}-shock. In both cases, assuming the synchrotron self-absorption frequency is well below the optical band, $\nu_a<\nu_{opt}$, the thickness of the shell and the density profile (``homogenous ISM'' [interstellar medium] or ``wind-like'' medium) affect the rate at which the lightcurve rises.

For instance, we are aware of no theoretical model that predicts such steep rising during the forward-shock, while if  the rising is due to the onset of the afterglow it is important to determine if we are in the ``thin-shell'' or ``thick-shell'' regime \citep[e.g.][]{Kobayashi:2000lq,Sari:1998vn}.
 For a ``thin-shell'', in a constant ISM, the temporal index is either $\alpha=-2$ ($\nu_c<\nu_{opt}$) or $\alpha=-3$ ($\nu_c >\nu_{opt}$, \cite{Zhang:2003dq}). A ``thick-shell'' would imply a much shallower rise in flux before the peak.
In case of a wind-like medium we expect a much shallower rise ($\alpha=-1/2$, \cite{Kobayashi:2003ul}).
Therefore is very unlikely that either a forward-shock emission or the passage of a synchrotron
characteristic frequency in the observed band could reproduce the observed early-time optical lightcurve.

A reverse-shock can produce the observed rising, but again it is critical to determine 
the regime in which the emission takes place \citep{Zhang:2003dq}. First of all, the $\alpha\approx -5$ can be only exist in a thin-shell regime, since the thick regime has a shallower rising $\alpha=-1/2$. 
We define $\Re_{\nu}= \frac{\nu_R}{\nu_{m,r}(t)}$, the ratio between the optical $R$ band frequency and the typical synchrotron frequency $\nu_m$. For the reverse shock,  $\Re_{\nu}>1 (\nu_{m,r}< \nu_R$, thin shell, reverse-shock). 
We can calculate $\nu_{m,r}$ and $\nu_{c,r}$ at the peak, 985 s after the burst, since we expect $\nu_{c,r}(t_{\rm peak})=\nu_{c,f}(t_{\rm peak})$, where the subscripts indicate the reverse and forward shock respectively.
From Eq. 1 of \cite{Zhang:2003dq}, we obtained that the critical Lorentz factor is $\gamma_c\approx550$, while the 
initial Lorentz factor can be estimated from the Lorentz factor at the peak time, $\gamma_{\rm peak}$ \citep{Meszaros:2006zr}:

\begin{center}
$
\gamma_0 = 2 \times \gamma_{\rm peak} = 2 \times \left( \frac{3(1+z)^3E}{32\pi nm_pc^5t_p^3}\right)^{1/8}  \approx 200 \left( \frac{E}{4\times 10^{53}~ \rm{erg}}\right)^{1/8}n^{-1/8}\left( \frac{1+z}{3.22}\right)^{3/8} \left( \frac{t_p}{985~ \rm{}s}\right)^{-3/8}
$.
\end{center}

For GRB 110205A we see that $\gamma_0< \gamma_c$, confirming that our assumption of thin shell regime is indeed correct.
At the time of the peak: $\nu_{c,r}=2.31 \times10^{16}$Hz (for $\epsilon_B=10^{-2}$) and $\nu_{m,f}=1.2 \times 10^{15}$. Finally, using $\nu_{m,r}\sim \gamma_0^{-2} \nu_{m,f}$ we can see that we are likely in $\nu_{m,r} < \nu_R < \nu_c \lesssim \nu_{X}$.  
Another point in favor of a thin-shell regime is the burst duration. In fact, in this case the deceleration time $t_{\gamma}$ is longer than the duration of the burst $T_{90}$, while for a thick-shell we would expect a much shorter time-scale.

If the optical band is located below the typical frequency of the \emph{forward}-shock at the deceleration time ($\nu_{opt}< \nu_{m,f}$),
the optical band pass is dominated by the \emph{reverse}-shock emission. Although this interpretation seems favorable
in explaining the temporal behavior, in this scenario the lightcurve should manifest the passage of $\nu_{m,f}$ in the observed bands
\citep[see Fig. 1 in][]{Kobayashi:2003ul}, and it is not consistent with the observations. 
Then, the optical band should be roughly around or above $\nu_{m,f}$ at the deceleration time. We can estimate this
from 
\begin{center}
$
\nu_{m,f} = (6\times 10^{15} Hz)(1+z)^{1/2}E_{52}^{1/2}\epsilon_e^2\epsilon_B^{1/2}(t/1 \,\rm{day})^{-3/2}. 
$
\end{center}

Which, for the observed value of $E_{iso}$, $z$, and time of the peak, and using typical values for $\epsilon_e$ and
 $\epsilon_B$, we estimate $\nu_{m,f} \approx 3\times 10^{14}$ Hz.
Under this condition, the onset of the afterglow is expected to be a single peak as observed in 
other cases \citep{Mundell:2007lr,Mundell:2007qy}. At the peak time, the 
contributions of the two shock emissions to the optical band are comparable, provided that the 
microscopic parameters are similar in the two shocked regions. The rapid rise 
is due to the bright reverse shock emission which masks the onset of the forward-shock emission, and it implies a weakly magnetized outflow from the central engine \citep{Fan:2002fr,Zhang:2003dq,
Kumar:2003zr,Gomboc:2008qy}.
 
\begin{figure*}
\epsscale{1.0}
\includegraphics[scale=0.8, angle=0]{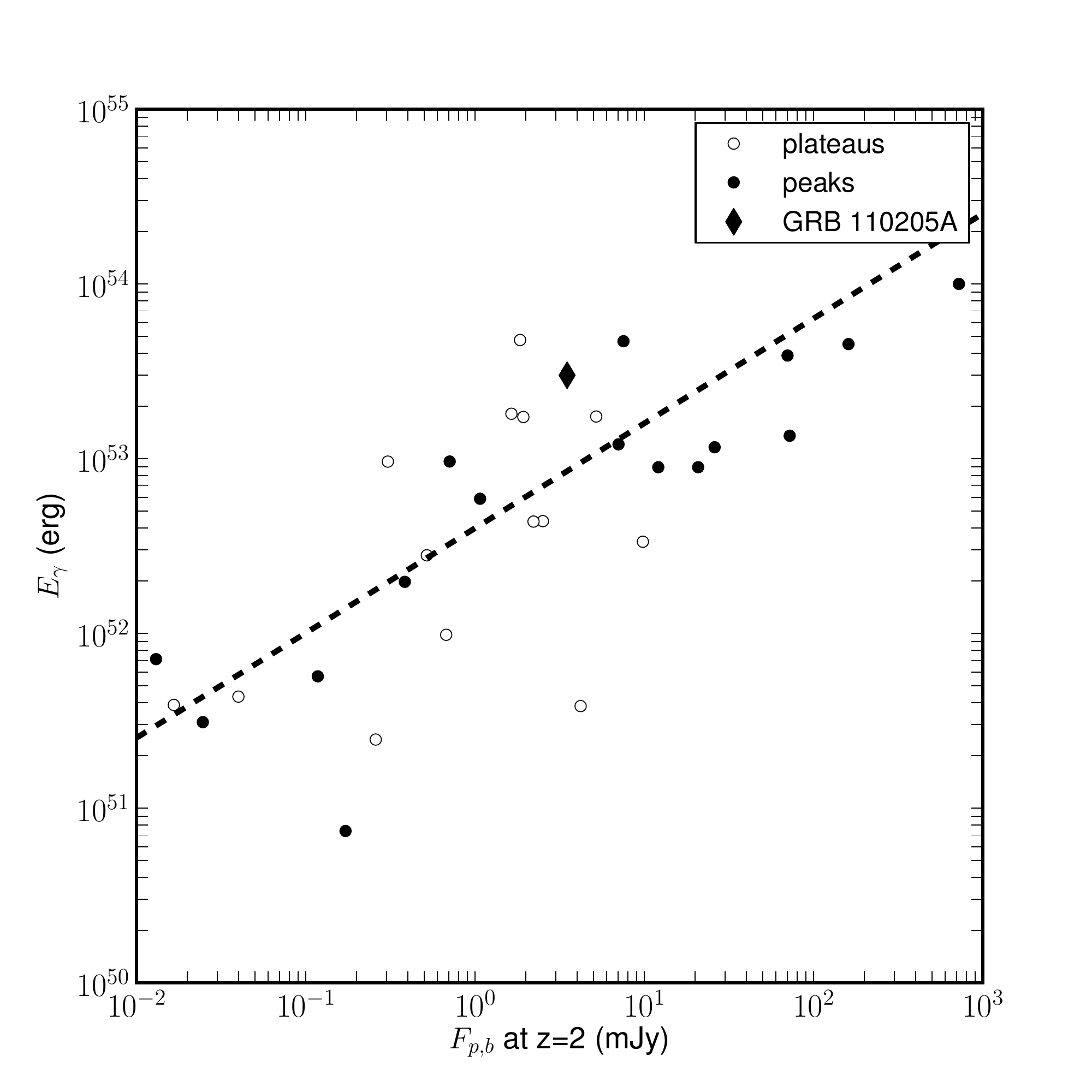}

\caption{Figure adopted from \cite{Panaitescu:2010bh}. The abscissa represents the optical flux at the observed peak time, $F_p$ (or at the end of an extended plateau phase, $T_b$). GRB 110205A is plotted as a filled diamond and presents a peak luminosity compared to other ``peaky'' afterglow cases (filled circles) or GRBs that present plateau phases (open circles). The dashed line represents the correlation between these fluxes and the isotropic equivalent energy released ($E_{\gamma}$). As mentioned by these authors the rising in the light curve may be due to a deceleration of the fireball which implies a large release of energy in a short amount of time. }
\label{fig:vestr}
\end{figure*}
\begin{figure*}
\epsscale{1.0}
\includegraphics[scale=0.50,angle= 0]{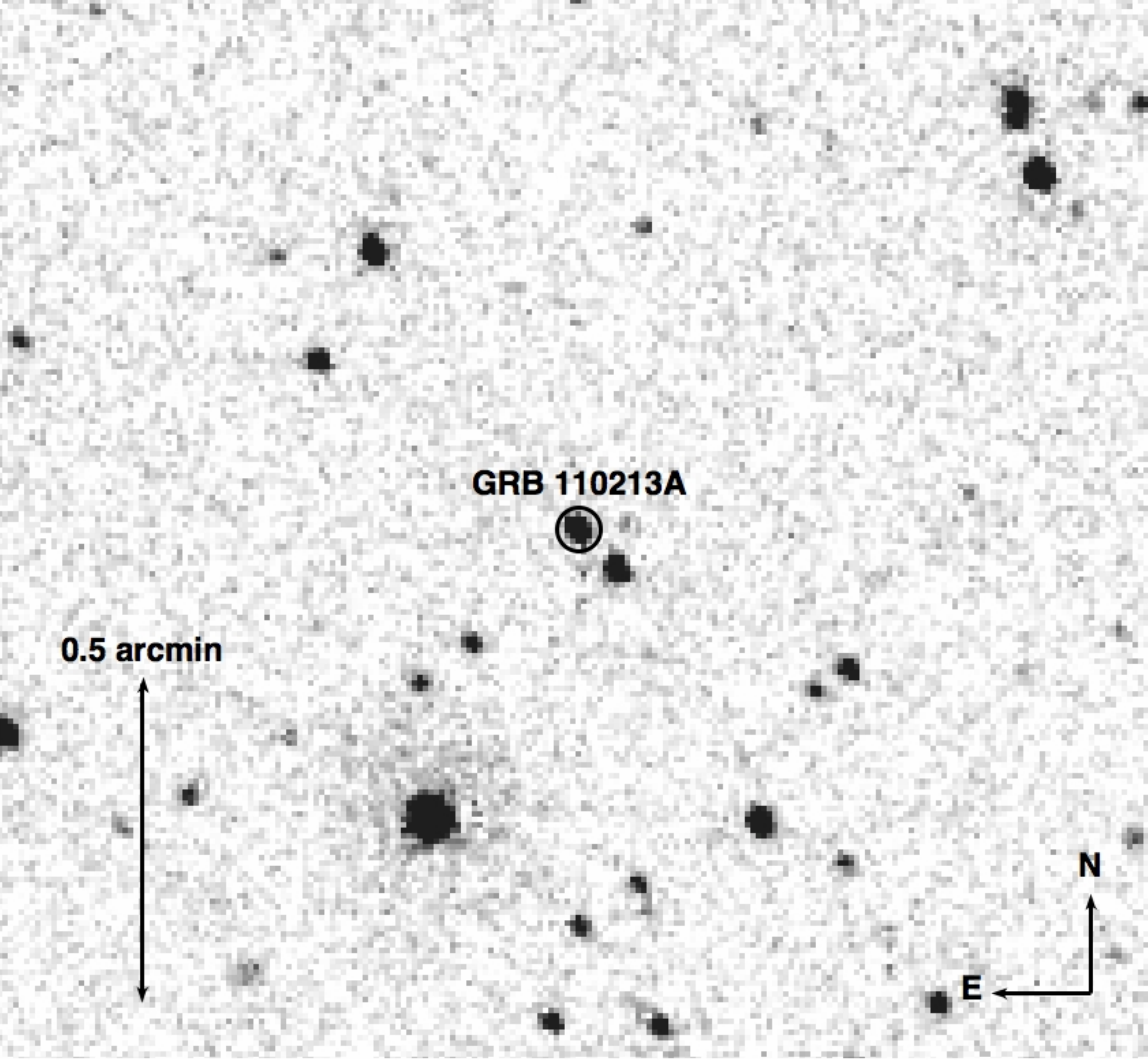}

\caption{UVOT $v$ band image of GRB 110213A obtained $\sim650$~s after the trigger.}
\label{fig:fin13A}
\end{figure*}
\begin{figure*}
\epsscale{1.0}
\includegraphics[scale=0.60,angle= 0]{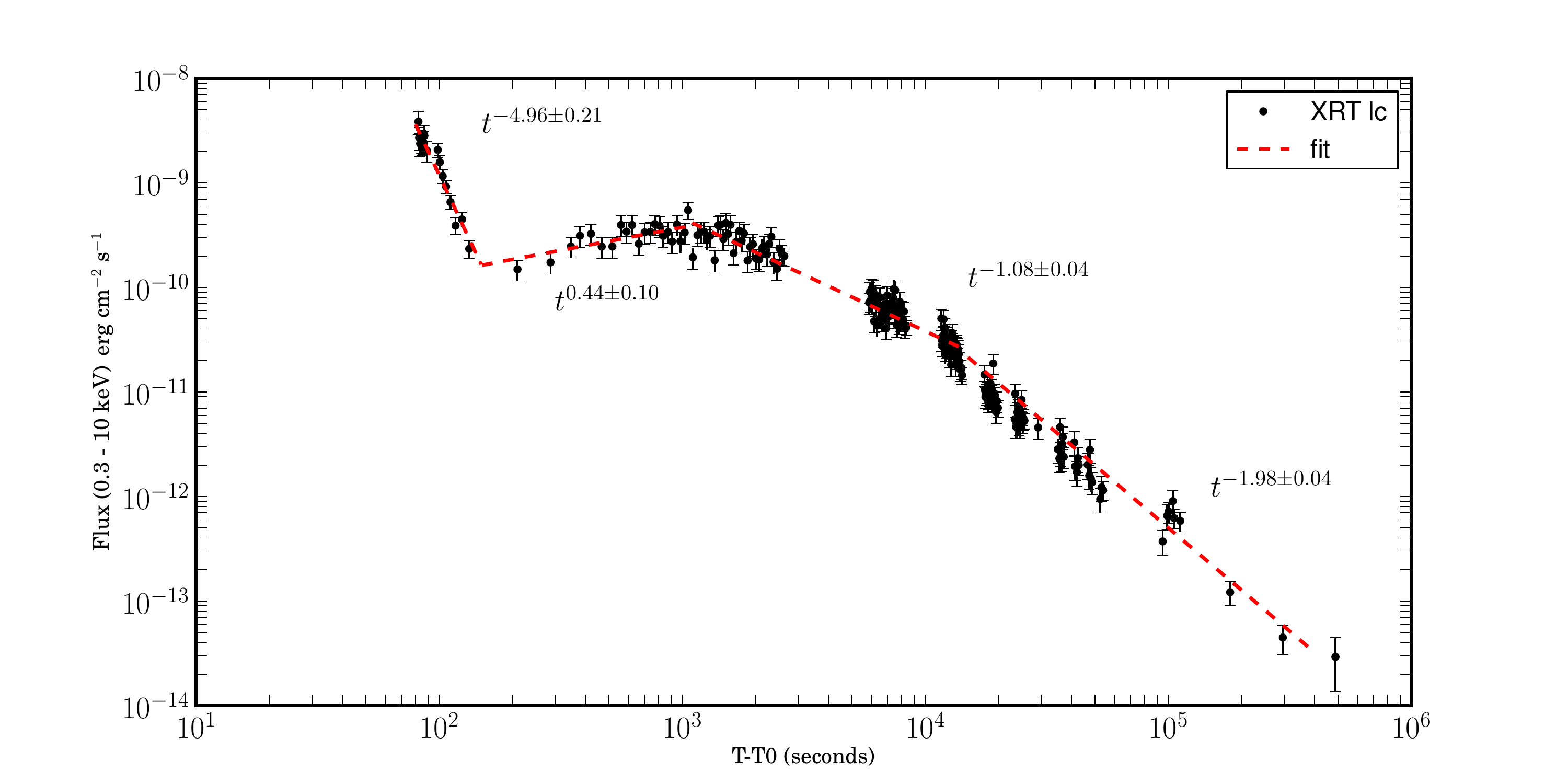}
\caption{GRB110213A XRT lightcurve. Dashed lines indicate the different power-law segments obtained fitting the XRT data with single power-laws.}
\label{fig:13Axrtlc}
\end{figure*}
\begin{figure*}

\epsscale{1.2}
\includegraphics[scale=0.50,angle=0]{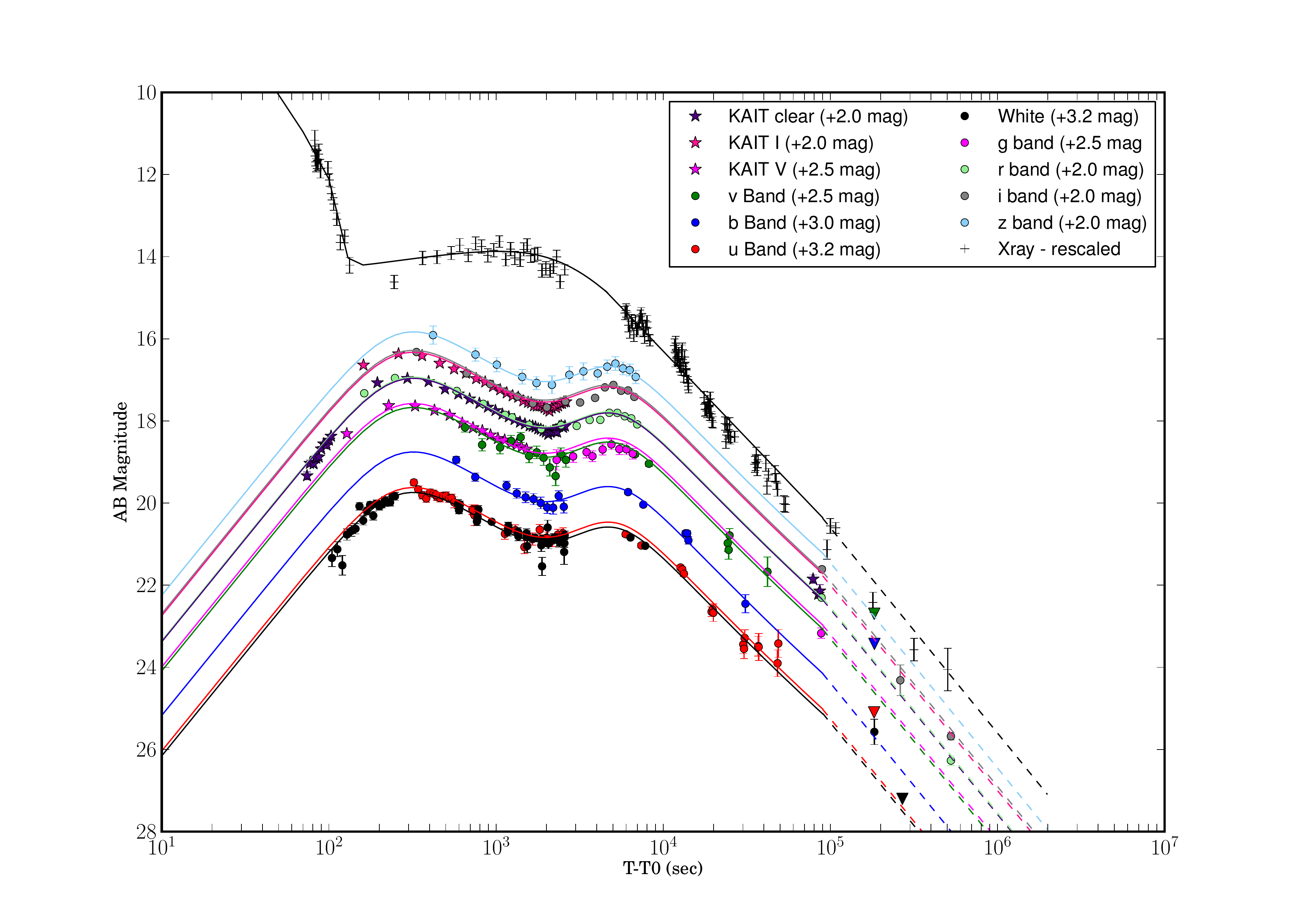}
\caption{GRB110213A lightcurve. We represent all the datasets present in this work. All of the magnitudes are transformed into AB systems \citep{Oke:1983uq}.
We interpret the two peaks as the onset of the afterglow and the continuous energy injection from a long living central engine. }
\label{fig:lc13A}
\end{figure*}
\begin{figure*}
\includegraphics[scale=0.6,angle=0]{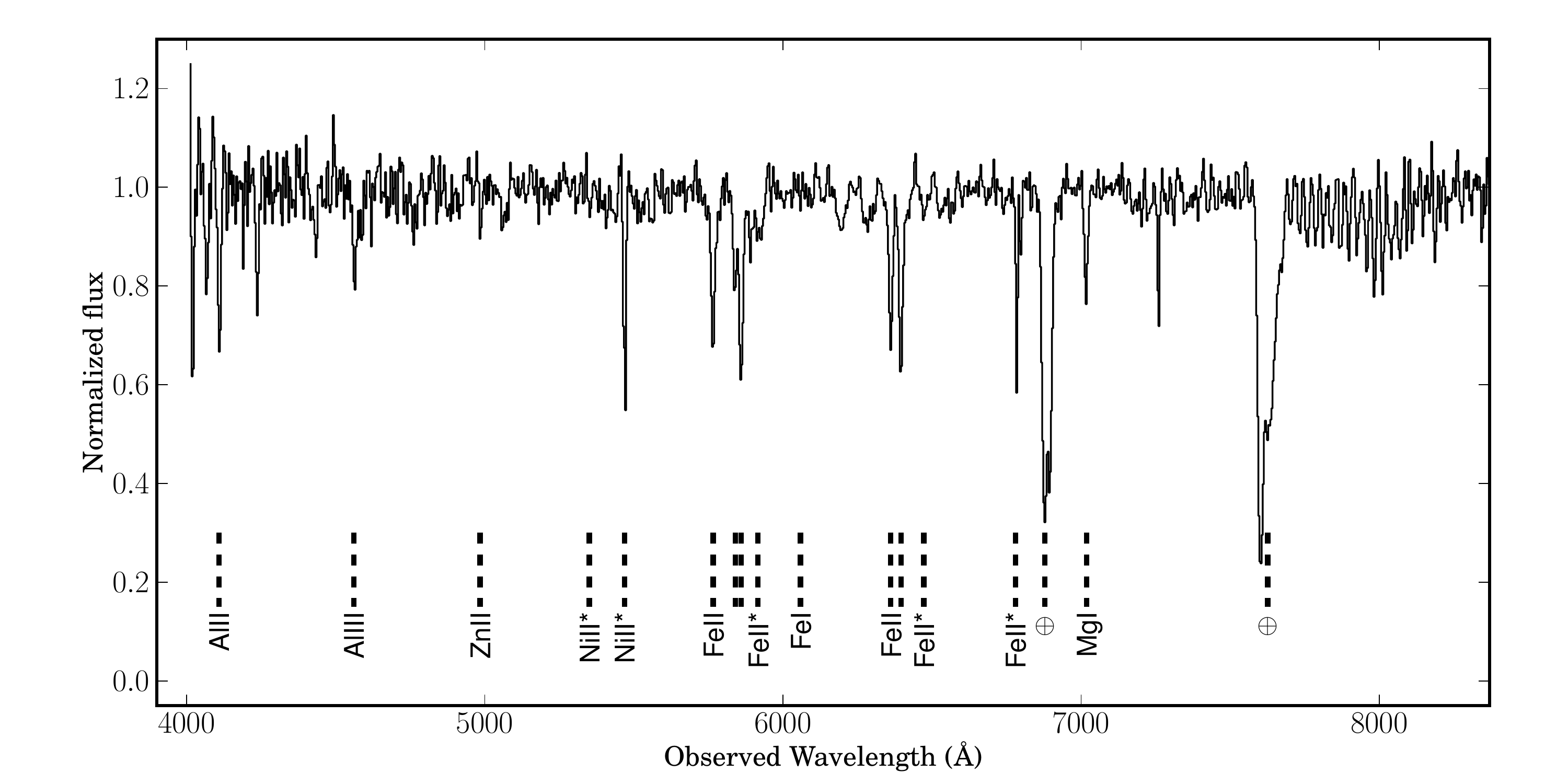}
\caption{Normalized spectrum of the afterglow of GRB 110213A obtained with the Bok spectrograph. The main absorption features are labeled, as are the telluric lines.} 
\label{fig:spec13A}
\end{figure*}

It is worthwhile to note that the peak flux density in V band is 3 mJy, as observed by UVOT, and it  is consistent with the sample of 
 \cite{Panaitescu:2010bh} \citep[see also][]{Vestrand:2006pd} for ``peaky'' afterglows, where, for a constant-ISM model and after a steep rise due to the fireball deceleration,
 the optical emission reaches a flux density of $\sim~ 4$ mJy (see Figure \ref{fig:vestr}).
After the onset of deceleration the afterglow is forward-shock dominated and evolves
 with the usual single power-law decay (except for the presence of a possible jet-break).


 \subsubsection{Late-time behavior}
After the reverse shock has passed through the GRB ejecta, the synchrotron emission 
produced by an external shock interacting with the ISM becomes the dominant emission mechanism.
The emission from the reverse shock decays as fast as $t^{-2}$,
so we expect a power-law decline as $t^{-\alpha}$, with $\alpha=(3p-2)/4$ (for $\nu_{opt}<\nu_c$) or 
$\alpha=3(p-1)/4$ (for $\nu_m < \nu_{opt}< \nu_c$). 
The late time optical-NIR decay indexes are $\alpha_{\rm opt}=1.74\pm0.28$, in agreement with the X-ray decay index $\alpha_{\rm X}=1.65\pm 0.05$, 
suggesting a forward shock producing the emission from the optical to the \xray. Assuming that the \xray\ afterglow emission
is dominated by forward shock, we estimated $p=2.90$ ($\alpha_{\rm X}=(3p-2)/4$), where $p$ is the index of the power-law distribution of random electrons accelerated at the shock \citep{Zhang:2003dq}. 

The observations performed by the RINGO2 polarimeter exclude the zero-polarization hypothesis
at 92\% confidence level, supporting the reverse plus forward shock scenario, in which the afterglow is mainly dominated by the forward shock. Nevertheless such low polarization as $P=3.6\%$ can be the result of several scenarios, like
a structured jet \citep{Rossi:2004mz}, the alignment of the magnetic field over causally connected regions 
\citep{Gruzinov:1999ly}, or even a large scale magnetic fields in the ambient medium \citep{Granot:2003gf}.
This result is also consistent with 
many other GRB polarization measurements \citep[see][and references within for the state of the art of polarization studies]{Covino:2004lr,Mundell:2010uq}.

\subsubsection{Absorption Spectrum}
We also observed the afterglow of GRB 110205A with the Kast
spectrograph on the 3~m Shane reflector at Lick Observatory.
Despite the fact that only a few absorption features  could be identified, a clear indication of high neutral hydrogen
in the host galaxy of this GRB comes from the detection of a DLA absorption system. 
A fit of the broad absorption profile indicates a value of log$(\nh/{\rm cm}^{-2}) = 21.45 \pm 0.2$ (Figure \ref{fig:DLA05A}), 
similar to the value obtained from the \xray\ data, perhaps indicating minimal photoionization of the surrounding medium
from the GRB itself \citep[see also][]{Campana:2010fk,Watson:2007ly}.


\subsection{GRB 110213A}
Optical data for GRB 110213A were obtained only after the prompt
emission had ended.  Nevertheless, our group was able to observe this
event from several different facilities, such as the P60 and KAIT,
providing much coverage of the afterglow phase.

\subsubsection{Prompt phase}
The prompt emission exhibits a single-peak profile and a soft spectrum. 
Using $E_{\rm peak}$ and $E_{\rm iso}$ derived by \swift\ and other
space facilities, GRB 110213A does not differ from 
other long-soft GRBs; thus, not surprisingly, it also obeys the 
Amati relation for long GRBs \citep{Amati:2008fy}.

\subsubsection{\xray\ lightcurve}
In the time between the beginning of the observations with XRT/UVOT and KAIT,
the \xray\ fades steeply with $\alpha_{\rm X}=4.96\pm0.21$.
This early steep declining phase, as for GRB 110205A, is consistent to be 
the tail of the prompt emission and it is governed by the curvature effect.
In particular, the spectral index and the temporal index are correlated by $\alpha= 2+\beta$. 
Using the estimated photon index from the first 150~s 
($\Gamma_{\rm X}=5.10\pm0.78$), we obtain $\alpha=6.10 \pm 0.78$, which is
consistent with the observed steep decay at a 1-$\sigma$ confident level.
At $T> 150$~s the \xray\ emission brightens with $\alpha_{\rm X}=-0.44\pm0.10$ up to 
$\sim1500$~s and then follows a shallow decay until $10^4$ s with $\alpha_{\rm X}=1.08\pm0.04$. 

Using  the spectral index  $\beta_{\rm X}=1.12\pm0.12$ we can determine that the behavior of 
GRB 110213A is consistent with the forward-shock scenario where the central engine is 
ejecting material in a slow-cooling regime in a homogenous ISM.
Using the closure relation $\alpha=(q-2)/2+(2+q)\beta/2$ we obtain  $\alpha=-0.44$,
where $q$ is defined as $q=2(\alpha+1-\beta)/(1+\beta)$, for $\nu>\nu_c,\nu_m$ 
and characterizes the central engine behavior \citep[where we have $q<1$ for an adiabatic fireball modified by
continued injection as presented in][]{Zhang:2006lr}.
The peak represents the cessation of the energy injection, after which the
normal adiabatic expansion of the fireball is expected. 
We derive for $10^4 \lesssim t \lesssim 10^6$~s a decay index of $\alpha_{\rm X} = 1.98 \pm 0.07$.
In summary, the overall \xray\ behavior of GRB 110213A is consistent with that of other GRBs in the \swift\
sample \citep[e.g.][]{Evans:2009uq}.

\subsubsection{Optical behavior - $T < 2000$ s}
KAIT observations of GRB 110213A indicate a rising afterglow with $\alpha= -2.08 \pm0.23$ from $T=70$ s until  $T=321$ s.
From the sample of \cite{Oates:2011fj,Oates:2009kx}, UVOT observed 6 GRBs with rising afterglows in the first 500~s.
GRB 110213A has a very similar case. 
What can be the cause of the peak at 321~s?
Several possibilities need to be tested: 1) the passage of one of the characteristic frequencies (e.g. $\nu_m$); 2) a reverse shock, as seen in GRB 110205A; 3) a decreasing extinction with time; 4) the onset of the forward shock.

The characteristic frequency $\nu_m$ that produces a peak in the lightcurve has a time dependence of $\nu_{m,f} \propto t^{-3/2}$ and should produce a chromatic break, meaning that the peak should be earlier in the bluer bands than the redder ones. 
During the rising part only KAIT, UVOT $white$ and 
P60 $r'$ band observations are available, so it is a challenging task to asses this chromaticity due to the gaps in the light curve. 
Nevertheless, during the passage of $\nu_m$ the spectral index changes from $\nu^{1/3}$ (for $\nu<\nu_m$) to $\nu^{-(p-1)/2}$ (for $\nu_m < \nu < \nu_c$). 
Based on this we can estimate that the color change between the UVOT $white$ filter and P60 $r'$
observation before and after the peak should be 0.68 mag. Instead, we measure a color difference of $0.15 \pm 0.10$ (after correcting for Galactic extinction). 
Furthermore, if we assume a constant density medium, $\nu_{m,f}$ can 
be calculated as:
\begin{center}
$
\nu_{m,f} = (6\times 10^{15} Hz)(1+z)^{1/2}E_{52}^{1/2}\epsilon_e^2\epsilon_B^{1/2}(t/1 \,\rm{day})^{-3/2}.
$
\end{center}
\noindent

For $z=1.46$, $E_{52}=7.2$ and $t=0.004$ day, and assuming that $\nu_{m,f}$ is already below the $r'$ band we obtain the constraint $\epsilon_e^2\epsilon_B^{1/2} < 5.6\times 10^{-6}$, consistent with other GRBs in the samples of \cite{Oates:2009kx} and  \cite{Panaitescu:2002fk}, in which the values of $\epsilon_e^2\epsilon_B^{1/2}$ range from $3\times 10^{-3}$ and $2 \times10^{-7}$.
Therefore we can exclude that the peak observed is due to the passage of $\nu_m$.
As suggested by some authors \citep{Klotz:2008yq,Rykoff:2004lr}, if there were a significant amount of extinction at the beginning of the afterglow phase the resulting light curve would be dim and reddened. Then, because of dust-destruction effect, it would rise faster in the blue filters than in the red ones. Of course this effect strongly depends on the GRB environment, but the KAIT, P60 and UVOT white observations, which all trace  the rising phase, show similar $\alpha$ index, from the blue bands (UVOT $white$) to the red (P60 $r'$). So we can also exclude that this effect is the dominant cause of the rising at early time.

The rising in the light curve at $T<321$~s is more likely due to the onset of the forward shock.
Unlike GRB 110205A, the rather slow rise does not require the domination of the reverse-shock 
emission at early times.  The post-peak decay $\alpha=1.10\pm0.24$ is also much shallower than 
the typical reverse-shock index ($\alpha=2$).  The forward-shock emission masks the 
reverse-shock emission when the typical frequency of the forward shock is below the optical band 
or when the magnetization of the ejecta is very high ($\sigma=B^2/4\pi\gamma \rho c^2$ is about unity or larger) and 
the magnetic pressure suppresses the reverse-shock emission (Giannios et al. 2008).  
The observed $\alpha=-2.08$ is consistent with  the expected $\alpha=-2$ for $\nu_c<\nu_{\rm opt}$, 
in a ``thin''-shell scenario for a homogeneous ISM.

Finally, for a constant density medium, we can use the peak time to determine the Lorentz factor at that time 
using the formalism of \cite{Sari:1999fj}, like in similar other cases \cite{Molinari:2007ys}. We obtain $\Gamma(t_{\rm peak})\approx139(\eta n_0)^{-1/8}$, where $\eta$ and $n_0$ are the radiative efficiency and the density of the shocked medium respectively. 
Usually the initial Lorentz factor is twice the value at peak, allowing us to estimate the deceleration radius of the fireball to be $R_{dec} \approx 2.27\times 10^{17}$ cm. The Lorentz factor and the deceleration radius are also consistent with the theoretical values predicted by \cite{Rees:1992rt}, as well as with the sample of \cite{Oates:2009kx}.

\subsubsection{Optical behavior - $T > 2000$ s}
A second peak at $t_{\rm peak}=4975\pm 545$~s is detected by several facilities, 
but is most prominent in data from the Palomar P60 and, 
in the decaying phase, the UVOT. Again, as previously, this feature can 
be due to one of the characteristic frequencies passing in the optical bands (in particular, 
at such late time, the synchrotron characteristic frequency $\nu_c$). 
Another possibility is that the emission is from the newly injected material in the blastwave 
(energy injection model).

In the case of synchrotron origin we calculated the optical spectral indices before and after 
the peak at $t_{peak}=4975$~s. 
We obtain $\beta_{o,pre}=1.12\pm0.24$ and $\beta_{o,post}=1.22\pm0.18$.
These values are consistent with each other within 1-$\sigma$, indicating that  $\nu_c < \nu_{z'}$ or $\nu_c> \nu_{white}$ at $t=t_{\rm peak}$.

Also, the  passage of a characteristic frequency during our observation would imply a 
chromatic feature, mainly a different peak time at different frequencies.
We tested this possibility assuming, for example, that the peak in $z'$ band at $t_{z',p}=4975$~s 
is due to the characteristic synchrotron frequency. Using the scaling law in \cite{Sari:1998vn}, we estimated that the peak would have crossed the UVOT $u$ band at: 

\begin{center}
$t_{u,p} = t_{z',p} \times (\frac{\nu_{c,z'}}{\nu_{c,u}})^2 = 743$ s.
\end{center}

\noindent From Figure \ref{fig:lc13A} it is evident that this is not the case.

One more possibility which would produce a ``bump'' in the lightcurve 
is the interaction of the fireball with moderate overdensity regions in the 
ambient medium \citep{Lazzati:2002ve}. Although, in this case, the 
lightcurve returns to the original power-law decay after the fireball
has passed through these overdensity regions, which is inconsistent with 
our observations.

Instead, the most likely scenario, in agreement with the \xray\ analysis, is that the 
rise and the following decays at late time are due to the re-injected material from the 
central engine into the ISM. At late time, in this case, we expect a
somewhat steeper decay than the usual adiabatic regime ($\alpha=(2-3p)/4$).
The decay index from the optical multiband fit is consistent with $\alpha=1.80 \pm 0.15$,
in agreement with the \xray\ emission. , whereas the adiabatic regime gives
$\alpha = 1.37$ assuming a typical value
of $p=2.5$.
Our assumption of a reinjection
phase better explains our late-time data.

Finally, using our Gemini ToO program we observed the
afterglow in the $r'$ and $i'$ bands around 6 days after the burst. The object is detected, but 
significantly below the extrapolation of the light curve from the early data.
This suggests the presence of a jet break after 1 day, placing a constraint on the 
opening angle of $\theta_{jet} \gtrsim 5^{\circ}$, and assuming the afterglow is the dominant source 
of radiation with a negligible contribution from the host galaxy.

\subsubsection{Absorption Spectrum}
The redshift of $z=1.46$, determined using the Bok telescope allows the identification
of UV rest-frame lines of low-ionization species. In particular, fine-structure transitions, like Ni~II* 
and Fe~II*, are indicative of UV pumping as the principle excitation mechanism in the vicinity of the GRB.
No neutral hydrogen estimate is possible, again, due to the low-redshift nature of this GRB.


\section{Conclusion}
\label{sec:discussion}
We observed GRB 110205A and GRB 110213A, discovered by the \swift\ satellite,  
with a broad range of follow-up facilities. The combination of our datasets and the publicly 
available \swift\ data covers more than six orders of magnitude in time.

GRB 110205A represents one of the best cases in which it has been possible to determine the
contribution of the \emph{reverse}-shock. While the UVOT and XRT data
trace very well the prompt phase, they also allow the characterization of a fast rising 
($\alpha=-6.13\pm0.75$) due to the reverse shock. After the peak at $T\approx985$~s,
the behavior is consistent with a combination of reverse and forward shocks, providing a 
shallower decay than the one expected for a pure $reverse$-shock.
A polarization measurement around 9 minutes after the burst provides only an upper limit 
(limiting polarization of $<16\%$),
with a 2-sigma linear polarization detection obtained around one hour post-burst ($P = 3.6 \%$).
This value, is in agreement with other estimates \citep[$P = 1.4 \%$, ][]{Gorosabel:2011uq}.
Unfortunately, in the absence of polarization variability information, we were unable to constrain 
the nature of the GRB environment or of the jet.
Nevertheless, the reverse shock emission dominates over 
the optical band and masks the forward shock emission at early times and it implies a weakly 
magnetized fireball.

Ground-based optical spectra reveal the presence of a strong DLA 
absorption system at 
$z=2.22$, produced by a neutral hydrogen column density of log$(\nh/{\rm cm}^{-2} = 21.45 \pm 0.2$.
This is not surprising, but it is interestingly similar to the 
value obtained from the \xray\ data, implying minimal photoionization of the surrounding medium
from the GRB itself.

In contrast, GRB 110213A presents a clear indication of a refreshed shock, most likely
produced by long-lived activity of the central engine. In the \xray, the plateau
phase is followed by the typical adiabatic behavior and by a steep decaying phase 
consistent with post-injection emission. In  the optical, using 7 bands (from the UVOT-$white$ to 
the P60 $z'$ filters), the afterglow presents two peaks due to the onset of the forward shock 
and to the interaction of the injected material with the ISM. Other interpretations, such as 
the passage of the characteristic synchrotron frequencies in the observed bands, are ruled out
since no chromatic features are present in our datasets. 
In the case of GRB 110213A, then, the forward-shock emission masks the 
reverse-shock emission, which means that  the typical frequency of the forward shock is lower than the
optical band or that the magnetic pressure is suppressing the reverse shock.

Using our late time Gemini observations we  were also able to detect the afterglow well beyond the 
capabilities of our small robotic telescopes. These observations provide some constraints
on the jet opening angle of the GRB emission. In particular we can place a lower limit of 
$\theta_{jet} \gtrsim 5^{\circ}$.

The importance of robotic facilities, multiband observations, and
spectroscopic follow-up reinforces the notion that a large array of facilities 
are needed in short order to interpret the 
complex early-time behavior of GRBs.
Future implementations, in particular using near-infrared cameras mounted on 
larger facilities, will allow the characterization of even higher redshift events, 
and will help test the dust properties of their environments \citep[e.g.][]{farah:77357Z}.
The synergy between ground-based and new-generation space-based observatories
will provide the best simultaneous coverage of these kind of events,
providing a complete description of the GRB phenomenon and GRB progenitors 
up to redshifts $z \gtrsim 9$.

\acknowledgements

AC acknowledges the anonymous referee for the precious comments
in order to improve the quality of this work.
The Gemini data, acquired under the program ID GN-20011A-Q-4, are
based on observations obtained at the Gemini Observatory, which is
operated by the Association of Universities for Research in Astronomy,
Inc., under a cooperative agreement with the NSF on behalf of the
Gemini partnership:  the National Science Foundation (United 
States), the Science and Technology Facilities Council (United Kingdom), the 
National Research Council (Canada), CONICYT (Chile), the Australian Research 
Council (Australia), Minist\'{e}rio da Ci\^{e}ncia e Tecnologia (Brazil) 
and Ministerio de Ciencia, Tecnolog\'{i}a e Innovaci\'{o}n Productiva (Argentina).
A.V.F., S.B.C., and W.L. acknowledge generous financial
assistance from Gary \& Cynthia Bengier, the Richard \& Rhoda Goldman Fund,
NASA/{\it Swift} grants NNX10AI21G and GO-7100028, the TABASGO Foundation,
and NSF grant AST-0908886. KAIT and its ongoing operation were made possible
by donations from Sun Microsystems, Inc., the Hewlett-Packard Company,
AutoScope Corporation, Lick Observatory, the NSF, the University of California,
the Sylvia \& Jim Katzman Foundation, and the TABASGO Foundation.
AC thanks C. Guidorzi and D.~A. Kann for the useful discussion and comments.


\bibliographystyle{apj_8}
\bibliography{bibi090429B,bibthesis,110205A,bib_master }


%
%
%
\clearpage
\begin{deluxetable}{lcccccc}
\tablewidth{0pc}
\tablecaption{Best Fit Temporal Parameters\label{tab:fits}}
\tabletypesize{\footnotesize}
\tablehead{\colhead{Event} & \colhead{$\alpha_{1}$} & \colhead{$\beta_{1}$} & \colhead{$t_{b,1}$} &\colhead{ $\alpha_{2}$} & \colhead{$\beta_{2}$} & \colhead{$t_{b,2}$}}
\startdata
GRB 110205A &$-6.13 \pm0.75$&$1.71\pm0.28$&$837^{+51}_{-40}$&$-0.48\pm0.22$&$1.74\pm0.28$&$68980$\\
GRB 110213A &$-2.08\pm0.23$&$1.10\pm0.24$&$263^{+13}_{-19}$&$-2.02\pm0.34$&$1.80\pm0.15$&$4827$\\
\enddata
\label{tab:fit05A}
\tablecomments{Best-fit parameters for GRB110205A and GRB110213A using the sum of two
 Beuermann function \citep{Beuermann:1999fk}. The multiband fit has been obtain simultaneously in all the observed optical and NIR
 bands.}
\end{deluxetable}

\clearpage
\begin{deluxetable}{ccccccccc}
\tablewidth{0pc}
\tablecaption{Absorption Lines in the Afterglow Spectrum of GRB~110205A\label{tab:ew}}
\tabletypesize{\footnotesize}
\tablehead{\colhead{$\lambda$} & \colhead{$z$} & \colhead{Transition} & \colhead{$W^a$} & \colhead{$\sigma(W)^b$} \\
(\AA) & & & (\AA) & (\AA) }
\startdata
4030.00 & 2.21420&SII 1253&$<1.10$&\\
4184.83 & 2.21374&OI 1302&$0.69$&0.23\\
4289.89 & 2.21453&CII 1334&$1.31$&0.25\\
4479.97 & 2.21432&SiIV 1393&$<1.09$&\\
4907.11 & 2.21418&SiII 1526&$0.99$&0.17\\
4927.55 & 2.21432&SiII* 1533&$0.54$&0.16\\
4977.05 & 2.21474&CIV 1548&$0.73$&0.20\\
5170.77 & 2.21408&FeII 1608&$0.86$&0.16\\
5371.76 & 2.21510&AlII 1670&$1.04$&0.15\\
5909.44 & &&$2.40$&0.59\\
5962.61 & 2.21484&AlIII 1854&$1.18$&0.19\\
6514.62 & 2.21529&ZnII 2026&$<0.51$&\\
\enddata
\tablenotetext{a}{Equivalent widths are rest-frame values and assume the redshift given in Column 2.}
\tablenotetext{b}{Uncertainties are $2 \sigma$ statistical values.}
\label{tab:05Alines}
\end{deluxetable}

\clearpage
\begin{deluxetable}{ccccccccc}
\tablewidth{0pc}
\tablecaption{Absorption Lines in the Afterglow Spectrum of GRB~110213A\label{tab:ew110213}}
\tabletypesize{\footnotesize}
\tablehead{\colhead{$\lambda$} & \colhead{$z$} & \colhead{Transition} & \colhead{$W^a$} & \colhead{$\sigma(W)^b$} \\
(\AA) & & & (\AA) & (\AA) }
\startdata
4112.10 & 1.46233&AlII 1670&$ 1.96$&0.33\\
4238.78 & &&$ 2.39$&0.68\\
4566.75 & 1.46311&AlIII 1854&$ 1.73$&0.21\\
4984.60 & 1.45934&ZnII 2026&$< 0.60$&\\
5310.17 & &&$ 2.77$&0.29\\
5354.79 & 1.46197&NiII* 2175&$ 1.16$&0.11\\
5404.37 & &&$ 3.12$&0.27\\
5433.04 & &&$ 2.27$&0.27\\
5471.65 & 1.46138&NiII* 2223&$ 1.82$&0.11\\
5767.82 & 1.46067&FeII 2344&$ 3.51$&0.09\\
5859.36 & &&$13.31$&0.22\\
5893.38 & 1.45967&FeII* 2396a&$ 4.54$&0.09\\
5918.62 & 1.46096&FeII* 2405&$ 6.89$&0.09\\
6237.86 & &&$80.12$&0.22\\
5998.19 & &&$ 6.16$&0.22\\
6062.10 & 1.46126&FeI 2463&$ 1.01$&0.09\\
6089.69 & &&$ 2.14$&0.21\\
6135.74 & &&$ 1.29$&0.21\\
6196.43 & &&$ 4.30$&0.21\\
6286.07 & &&$ 5.45$&0.21\\
6363.63 & 1.46080&FeII 2586&$ 2.73$&0.08\\
6397.78 & 1.46068&FeII 2600&$ 3.31$&0.08\\
6475.33 & 1.46116&FeII* 2631&$ 1.76$&0.08\\
6530.68 & &&$ 3.51$&0.20\\
6562.87 & &&$ 4.03$&0.20\\
6786.08 & 1.46229&FeII* 2756&$ 1.20$&0.07\\
7018.68 & 1.46010 & MgI 2853& $1.81$&0.19\\ 
\enddata
\tablenotetext{a}{Equivalent widths are rest-frame values and assume the redshift given in Column 2.}
\tablenotetext{b}{Uncertainties are $2 \sigma$ statistical values.}
\label{tab:0213Alines}
\end{deluxetable}


%
%
%
\clearpage

\clearpage


%

\clearpage


%

\begin{deluxetable}{lccc} 
\tablecolumns{4} 
\tablewidth{0pc}
\tablecaption{Photometric observations for GRB 110205A} 
\tabletypesize{\footnotesize}
\tablehead{ 
\colhead{$T-T_0$}  &   \colhead{Filter} &  \colhead{Magnitude$^a$} &
\colhead{1$\sigma$ Error}\\
\,\,\,\,(s) &&}
\startdata 
\multicolumn{4}{c}{\swift-UVOT}\\
\cline{1-4}\\
166  & $white$ & 18.68 & 0.34   \\  
171  & $white$ & 18.86 & 0.38   \\  
176  & $white$ & 18.79 & 0.38   \\  
186  & $white$ & 18.59 & 0.31   \\  
201  & $white$ & 18.67 & 0.35   \\  
206  & $white$ & 18.52 & 0.31   \\  
211  & $white$ & 17.47 & 0.17   \\  
216  & $white$ & 17.77 & 0.19   \\  
221  & $white$ & 18.15 & 0.25   \\  
226  & $white$ & 18.48 & 0.29   \\  
231  & $white$ & 18.45 & 0.30   \\  
241  & $white$ & 18.59 & 0.32   \\  
266  & $white$ & 18.75 & 0.36   \\  
286  & $white$ & 18.63 & 0.33   \\  
291  & $white$ & 18.73 & 0.34   \\  
301  & $white$ & 18.73 & 0.36   \\  
601  & $white$ & 16.74 & 0.18   \\  
606  & $white$ & 17.24 & 0.15   \\  
611  & $white$ & 17.03 & 0.13   \\  
616  & $white$ & 16.95 & 0.13   \\  
621  & $white$ & 16.89 & 0.17   \\  
776  & $white$ & 15.73 & 0.09   \\  
781  & $white$ & 15.78 & 0.08   \\  
786  & $white$ & 15.70 & 0.08   \\  
791  & $white$ & 15.80 & 0.08   \\  
1181  & $white$ & 15.25 & 0.09   \\  
1186  & $white$ & 15.20 & 0.07   \\  
1191  & $white$ & 15.40 & 0.07   \\  
1196  & $white$ & 15.27 & 0.07   \\  
1201  & $white$ & 15.37 & 0.12   \\  
1356  & $white$ & 15.45 & 0.09   \\  
1361  & $white$ & 15.58 & 0.08   \\  
1366  & $white$ & 15.55 & 0.08   \\  
1371  & $white$ & 15.55 & 0.08   \\  
1376  & $white$ & 15.25 & 0.12   \\  
1531  & $white$ & 15.73 & 0.08   \\  
1536  & $white$ & 15.76 & 0.08   \\  
1541  & $white$ & 15.76 & 0.08   \\  
1546  & $white$ & 15.65 & 0.08   \\  
1701  & $white$ & 15.78 & 0.12   \\  
1706  & $white$ & 15.65 & 0.08   \\  
1711  & $white$ & 16.20 & 0.09   \\  
1716  & $white$ & 15.96 & 0.09   \\  
1721  & $white$ & 16.13 & 0.13   \\  
5166  & $white$ & 17.67 & 0.04   \\  
6602  & $white$ & 18.07 & 0.04   \\  
11906  & $white$ & 19.15 & 0.06   \\  
12210  & $white$ & 19.16 & 0.06   \\  
12513  & $white$ & 19.15 & 0.06   \\  
24517  & $white$ & 20.04 & 0.12   \\  
24734  & $white$ & 20.25 & 0.22   \\  
35676  & $white$ & 20.69 & 0.21   \\  
35979  & $white$ & 20.70 & 0.21   \\  
36284  & $white$ & 21.28 & 0.35   \\  
81699  & $white$ & 21.93 & 0.32   \\  
134282 & $white$ &$>23.79$ &\\
\enddata 
\label{tab:tab1}
\tablenotetext{a}{UVOT magnitude are in the natural UVOT photometric system (Vega). See \cite{Breeveld:2011kx} for the most updated zeropoint and conversion factors.}

\end{deluxetable}

\begin{deluxetable}{lccc} 
\tablecolumns{4} 
\tablewidth{0pc}
\tablecaption{Photometric observations for GRB 110205A} 
\tabletypesize{\footnotesize}
\tablehead{ 
\colhead{$T-T_0$}  &   \colhead{Filter} &  \colhead{Magnitude$^a$} &
\colhead{1$\sigma$ Error}\\
\,\,\,\,(s) &&}
\startdata 
\multicolumn{4}{c}{\swift-UVOT}\\
\cline{1-4}\\
409  & $u$ & 18.75 & 0.39   \\  
434  & $u$ & 18.71 & 0.37   \\  
459  & $u$ & 18.58 & 0.35   \\  
484  & $u$ & 18.33 & 0.30   \\  
509  & $u$ & 17.92 & 0.22   \\  
534  & $u$ & 17.32 & 0.16   \\  
559  & $u$ & 17.06 & 0.14   \\  
734  & $u$ & 15.58 & 0.08   \\  
1134  & $u$ & 14.97 & 0.07   \\  
1159  & $u$ & 14.88 & 0.14   \\  
1309  & $u$ & 15.15 & 0.07   \\  
1334  & $u$ & 14.98 & 0.15   \\  
1484  & $u$ & 15.30 & 0.07   \\  
1659  & $u$ & 15.53 & 0.08   \\  
1834  & $u$ & 15.67 & 0.09   \\  
47582  & $u$ & 20.65 & 0.48   \\  
67156  & $u$ & 21.01 & 0.22   \\
133453  & $u$ & $>23.31$&\\
\enddata 
\tablenotetext{a}{UVOT magnitude are in the natural UVOT photometric system (Vega). See \cite{Breeveld:2011kx} for the most updated zeropoint and conversion factors.}
\label{tab:tab2}
\end{deluxetable}

\begin{deluxetable}{lccc} 
\tablecolumns{4} 
\tablewidth{0pc}
\tablecaption{Photometric observations for GRB 110205A} 
\tabletypesize{\footnotesize}
\tablehead{ 
\colhead{$T-T_0$}  &   \colhead{Filter} &  \colhead{Magnitude$^a$} &
\colhead{1$\sigma$ Error}\\
\,\,\,\,(s) &&}\startdata 
\multicolumn{4}{c}{\swift-UVOT}\\
\cline{1-4}\\
587  & $b$ & 17.08 & 0.12   \\  
747  & $b$ & 15.80 & 0.10   \\  
767  & $b$ & 15.77 & 0.08   \\  
1147  & $b$ & 15.32 & 0.13   \\  
1167  & $b$ & 15.22 & 0.06   \\  
1327  & $b$ & 15.41 & 0.10   \\  
1347  & $b$ & 15.47 & 0.07   \\  
1507  & $b$ & 15.77 & 0.08   \\  
1527  & $b$ & 15.72 & 0.11   \\  
1687  & $b$ & 15.92 & 0.07   \\  
1847  & $b$ & 16.01 & 0.12   \\  
1867  & $b$ & 16.11 & 0.09   \\  
6397  & $b$ & 17.99 & 0.06   \\  
10993  & $b$ & 19.10 & 0.11   \\  
11297  & $b$ & 19.00 & 0.10   \\  
11601  & $b$ & 18.97 & 0.10   \\  
23604  & $b$ & 20.10 & 0.23   \\  
23908  & $b$ & 20.10 & 0.24   \\  
24212  & $b$ & 20.45 & 0.31   \\  
30449  & $b$ & 20.38 & 0.30   \\  
34763  & $b$ & 20.36 & 0.28   \\  
61506  & $b$ & 21.75 & 0.31   \\  
133867& $b$ &$>22.63$& \\
\enddata 
\tablenotetext{a}{UVOT magnitude are in the natural UVOT photometric system (Vega). See \cite{Breeveld:2011kx} for the most updated zeropoint and conversion factors.}

\label{tab:tab3}
\end{deluxetable}

\begin{deluxetable}{lccc} 
\tablecolumns{4} 
\tablewidth{0pc}
\tablecaption{Photometric observations for GRB 110205A} 
\tabletypesize{\footnotesize}
\tablehead{ 
\colhead{$T-T_0$}  &   \colhead{Filter} &  \colhead{Magnitude$^a$} &
\colhead{1$\sigma$ Error}\\
\,\,\,\,(s) &&}
\startdata 
\multicolumn{4}{c}{\swift-UVOT}\\
\cline{1-4}\\
655  & $v$ & 15.88 & 0.14   \\  
675  & $v$ & 15.63 & 0.18   \\  
835  & $v$ & 15.04 & 0.08   \\  
1055  & $v$ & 14.62 & 0.09   \\  
1075  & $v$ & 14.65 & 0.09   \\  
1235  & $v$ & 14.65 & 0.08   \\  
1255  & $v$ & 14.80 & 0.12   \\  
1415  & $v$ & 15.19 & 0.08   \\  
1575  & $v$ & 15.10 & 0.13   \\  
1595  & $v$ & 15.26 & 0.10   \\  
1755  & $v$ & 15.24 & 0.10   \\  
1775  & $v$ & 15.52 & 0.17   \\  
5576  & $v$ & 17.34 & 0.08   \\  
7013  & $v$ & 17.79 & 0.11   \\  
16772  & $v$ & 19.04 & 0.22   \\  
17076  & $v$ & 19.17 & 0.25   \\  
17379  & $v$ & 19.10 & 0.23   \\  
134686 & $v$ & $>22.82$&    \\  
\enddata 
\tablenotetext{a}{UVOT magnitude are in the natural UVOT photometric system (Vega). See \cite{Breeveld:2011kx} for the most updated zeropoint and conversion factors.}

\label{tab:tab4}
\end{deluxetable}

\begin{deluxetable}{lccc} 
\tablecolumns{4} 
\tablewidth{0pc}
\tablecaption{Photometric observations for GRB 110205A} 
\tabletypesize{\footnotesize}
\tablehead{ 
\colhead{$T-T_0$}  &   \colhead{Filter} &  \colhead{Magnitude} &
\colhead{1$\sigma$ Error}\\
\,\,\,\,(s) &&}
\startdata 
\multicolumn{4}{c}{Palomar telescope}\\
\cline{1-4}\\
7929  & \gp & 18.272 & 0.162   \\  
18421  & \gp & 19.265 & 0.036   \\  
19229  & \gp & 19.421 & 0.028   \\  
20026  & \gp & 19.485 & 0.013   \\  
20913  & \gp & 19.522 & 0.019   \\  
21730  & \gp & 19.627 & 0.019   \\  
22533  & \gp & 19.659 & 0.019   \\  
23350  & \gp & 19.746 & 0.019   \\  
24158  & \gp & 19.819 & 0.022   \\  
24961  & \gp & 19.819 & 0.017   \\  
25763  & \gp & 19.883 & 0.039   \\  
26561  & \gp & 19.964 & 0.032   \\  
27358  & \gp & 20.072 & 0.027   \\  
28175  & \gp & 20.02 & 0.026   \\  
28984  & \gp & 20.109 & 0.022   \\  
29886  & \gp & 20.168 & 0.024   \\  
30704  & \gp & 20.235 & 0.024   \\  
31531  & \gp & 20.238 & 0.041   \\  
32334  & \gp & 20.25 & 0.03   \\  
33131  & \gp & 20.333 & 0.026   \\  
33934  & \gp & 20.444 & 0.032   \\  
34741  & \gp & 20.427 & 0.042   \\  
35538  & \gp & 20.495 & 0.048   \\  
36331  & \gp & 20.599 & 0.058   \\  
37128  & \gp & 20.714 & 0.032   \\  
37921  & \gp & 20.69 & 0.042   \\  
38714  & \gp & 20.795 & 0.06   \\  
39579  & \gp & 20.873 & 0.06   \\  
40372  & \gp & 20.788 & 0.061   \\  
41164  & \gp & 20.891 & 0.093   \\  
\cline{1-4}\\
\multicolumn{4}{c}{Liverpool Telescope}\\
\cline{1-4}\\
4221 & \gp & 17.266 & 0.109\\
4423 & \gp & 17.286 & 0.039\\
4630 & \gp & 17.226 & 0.109\\
4926 & \gp & 17.356 & 0.109\\
5223 & \gp & 17.506 & 0.109\\
6103 & \gp & 17.606 & 0.109\\
6393 & \gp & 17.776 & 0.109\\
6500 & \gp & 17.756 & 0.109\\
6976 & \gp & 17.926 & 0.109\\
7631 & \gp & 18.126 & 0.109\\
8105 & \gp & 18.156 & 0.109\\
8763 & \gp & 18.336 & 0.109\\
9239 & \gp & 18.386 & 0.109\\
\enddata 
\label{tab:tab5}
\end{deluxetable}

\begin{deluxetable}{lccc} 
\tablecolumns{4} 
\tablewidth{0pc}
\tablecaption{Photometric observations for GRB 110205A} 
\tabletypesize{\footnotesize}
\tablehead{ 
\colhead{$T-T_0$}  &   \colhead{Filter} &  \colhead{Magnitude} &
\colhead{1$\sigma$ Error}\\
\,\,\,\,(s) &&}
\startdata 
\multicolumn{4}{c}{KAIT telescope}\\
\cline{1-4}\\
14344  & \rp & 18.499 & 0.200   \\  
14470  & \rp & 18.577 & 0.217   \\  
14594  & \rp & 18.543 & 0.225   \\  
14719  & \rp & 18.540 & 0.194   \\  
14844  & \rp & 18.608 & 0.200   \\  
14968  & \rp & 18.518 & 0.182   \\  
15093  & \rp & 18.588 & 0.208   \\  
15218  & \rp & 18.632 & 0.190   \\  
15343  & \rp & 18.686 & 0.205   \\  
15468  & \rp & 18.535 & 0.199   \\  
15593  & \rp & 18.670 & 0.208   \\  
15717  & \rp & 18.769 & 0.216   \\  
15842  & \rp & 18.735 & 0.204   \\  
15965  & \rp & 18.899 & 0.221   \\  
16090  & \rp & 18.944 & 0.206   \\  
16214  & \rp & 18.720 & 0.189   \\  
16338  & \rp & 18.797 & 0.219   \\  
16462  & \rp & 18.733 & 0.208   \\  
16587  & \rp & 18.863 & 0.211   \\  
16712  & \rp & 18.778 & 0.198   \\  
17781  & \rp & 18.863 & 0.198   \\  
17905  & \rp & 18.980 & 0.211   \\  
18028  & \rp & 18.873 & 0.193   \\  
18153  & \rp & 18.848 & 0.195   \\  
18278  & \rp & 18.921 & 0.182   \\  
18400  & \rp & 18.962 & 0.189   \\  
18523  & \rp & 18.875 & 0.189   \\  
18647  & \rp & 18.889 & 0.196   \\  
18772  & \rp & 18.917 & 0.194   \\  
18897  & \rp & 19.045 & 0.219   \\  
19021  & \rp & 18.929 & 0.202   \\  
19146  & \rp & 18.860 & 0.191   \\  
20127  & \rp & 18.974 & 0.148   \\  
20408  & \rp & 18.852 & 0.147   \\  
20686  & \rp & 19.010 & 0.145   \\  
20970  & \rp & 19.081 & 0.139   \\  
21254  & \rp & 19.072 & 0.146   \\  
22514  & \rp & 19.228 & 0.136   \\  
30129  & \rp & 19.690 & 0.136   \\  
31342  & \rp & 19.687 & 0.142   \\  
34080  & \rp & 19.816 & 0.126   \\  
\cline{1-4}\\
\multicolumn{4}{c}{Palomar Telescope}\\
\cline{1-4}\\
5766  & \rp & 17.090 & 0.018   \\  
5969  & \rp & 17.126 & 0.013   \\  
6172  & \rp & 17.180 & 0.015   \\  
7528  & \rp & 17.533 & 0.134   \\  
14775  & \rp & 18.526 & 0.045   \\  
18823  & \rp & 18.935 & 0.019   \\  
19625  & \rp & 19.000 & 0.018   \\  
20428  & \rp & 19.069 & 0.015   \\  
21324  & \rp & 19.119 & 0.017   \\  
22127  & \rp & 19.159 & 0.024   \\  
23752  & \rp & 19.335 & 0.026   \\  
24554  & \rp & 19.343 & 0.028   \\  
25357  & \rp & 19.403 & 0.039   \\  
26164  & \rp & 19.464 & 0.041   \\  
26957  & \rp & 19.514 & 0.036   \\  
27764  & \rp & 19.561 & 0.024   \\  
28578  & \rp & 19.646 & 0.035   \\  
29390  & \rp & 19.735 & 0.019   \\  
30303  & \rp & 19.797 & 0.030   \\  
31125  & \rp & 19.692 & 0.036   \\  
31928  & \rp & 19.800 & 0.022   \\  
32730  & \rp & 19.779 & 0.033   \\  
33528  & \rp & 19.889 & 0.039   \\  
34331  & \rp & 19.969 & 0.081   \\  
35143  & \rp & 20.029 & 0.028   \\  
35935  & \rp & 20.050 & 0.030   \\  
36733  & \rp & 20.147 & 0.042   \\  
37525  & \rp & 20.177 & 0.036   \\  
38317  & \rp & 20.161 & 0.045   \\  
39183  & \rp & 20.351 & 0.055   \\  
39976  & \rp & 20.306 & 0.038   \\  
40768  & \rp & 20.288 & 0.093   \\  
41561  & \rp & 20.369 & 0.138   \\  
91963  & \rp & 22.244 & 0.153   \\ 
\cline{1-4}\\
\multicolumn{4}{c}{Liverpool Telescope}\\
\cline{1-4}\\
540 & \rp & 16.92 & 0.68 \\
922.2 & \rp & 14.518 & 0.011 \\
945 & \rp & 14.508 & 0.011 \\
967.8 & \rp & 14.438 & 0.013 \\
1015 & \rp & 14.418 & 0.013 \\
1037 & \rp & 14.418 & 0.012 \\
1059 & \rp & 14.318 & 0.015 \\
3360 & \rp & 16.37 & 0.07 \\
3708 & \rp & 16.458 & 0.046 \\
3730 & \rp & 16.448 & 0.048 \\
3752 & \rp & 16.458 & 0.059 \\
3798 & \rp & 16.558 & 0.084 \\
3821 & \rp & 16.518 & 0.171 \\
3843 & \rp & 16.688 & 0.101 \\
4355 & \rp & 16.588 & 0.102 \\
4559 & \rp & 16.708 & 0.012 \\
4827 & \rp & 16.828 & 0.007 \\
5124 & \rp & 16.928 & 0.027 \\
6003 & \rp & 17.378 & 0.258 \\
6818 & \rp & 17.418 & 0.05 \\
7413 & \rp & 17.568 & 0.012 \\
7947 & \rp & 17.678 & 0.009 \\
8544 & \rp & 17.778 & 0.014 \\
9081 & \rp & 17.878 & 0.009 \\
7.029e+04 & \rp & 21.208 & 0.086 \\
7.06e+04 & \rp & 21.368 & 0.103 \\
\cline{1-4}\\
\multicolumn{4}{c}{Gemini-N Telescope}\\
\cline{1-4}\\
384191  & \rp & 23.450 & 0.050   \\  
2937600 & \rp & $>27.21$ & \\
\enddata 
\label{tab:tab6}
\end{deluxetable}

\begin{deluxetable}{lccc} 
\tablecolumns{4} 
\tablewidth{0pc}
\tablecaption{Photometric observations for GRB 110205A} 
\tabletypesize{\footnotesize}
\tablehead{ 
\colhead{$T-T_0$}  &   \colhead{Filter} &  \colhead{Magnitude} &
\colhead{1$\sigma$ Error}\\
\,\,\,\,(s) &&}
\startdata 
\multicolumn{4}{c}{Palomar telescope}\\
\cline{1-4}\\
6375  & \ip & 16.945 & 0.019   \\  
6574  & \ip & 16.992 & 0.015   \\  
6772  & \ip & 17.057 & 0.018   \\  
7731  & \ip & 17.368 & 0.161   \\  
14973  & \ip & 18.376 & 0.045   \\  
18214  & \ip & 18.646 & 0.09   \\  
19031  & \ip & 18.689 & 0.017   \\  
19828  & \ip & 18.756 & 0.013   \\  
20710  & \ip & 18.833 & 0.018   \\  
21527  & \ip & 18.89 & 0.013   \\  
22330  & \ip & 18.957 & 0.03   \\  
23152  & \ip & 18.984 & 0.024   \\  
23955  & \ip & 19.095 & 0.019   \\  
24757  & \ip & 19.164 & 0.019   \\  
25560  & \ip & 19.22 & 0.028   \\  
26362  & \ip & 19.241 & 0.033   \\  
27160  & \ip & 19.341 & 0.022   \\  
27972  & \ip & 19.444 & 0.06   \\  
28776  & \ip & 19.353 & 0.045   \\  
29593  & \ip & 19.417 & 0.024   \\  
30500  & \ip & 19.438 & 0.038   \\  
31333  & \ip & 19.522 & 0.042   \\  
32131  & \ip & 19.524 & 0.039   \\  
32928  & \ip & 19.566 & 0.036   \\  
33731  & \ip & 19.675 & 0.044   \\  
34538  & \ip & 19.658 & 0.075   \\  
35341  & \ip & 19.724 & 0.033   \\  
36133  & \ip & 19.851 & 0.048   \\  
36930  & \ip & 19.838 & 0.057   \\  
37723  & \ip & 19.876 & 0.048   \\  
38515  & \ip & 20.021 & 0.048   \\  
39382  & \ip & 20.027 & 0.078   \\  
40174  & \ip & 20.057 & 0.038   \\  
40966  & \ip & 20.132 & 0.078   \\  
92572  & \ip & 22.058 & 0.213   \\  
386587  & \ip & 23.45 & 0.05   \\  
\cline{1-4}\\
\multicolumn{4}{c}{Liverpool Telescope}\\
\cline{1-4}\\
4289 & \ip & 16.477 & 0.093 \\
4491 & \ip & 16.547 & 0.016 \\
4729 & \ip & 16.547 & 0.008 \\
5028 & \ip & 16.627 & 0.032 \\
5906 & \ip & 16.687 & 0.345 \\
6660 & \ip & 17.087 & 0.035 \\
7196 & \ip & 17.177 & 0.013 \\
7789 & \ip & 17.317 & 0.044 \\
8326 & \ip & 17.507 & 0.013 \\
8923 & \ip & 17.577 & 0.013 \\
\enddata 
\label{tab:tab7}
\end{deluxetable}

\begin{deluxetable}{lccc} 
\tablecolumns{4} 
\tablewidth{0pc}
\tablecaption{Photometric observations for GRB 110205A} 
\tabletypesize{\footnotesize}
\tablehead{ 
\colhead{$T-T_0$}  &   \colhead{Filter} &  \colhead{Magnitude} &
\colhead{1$\sigma$ Error}\\
\,\,\,\,(s) &&}
\startdata 
\multicolumn{4}{c}{Palomar telescope}\\
\cline{1-4}\\
18620  & \zp & 18.485 & 0.081   \\  
19427  & \zp & 18.352 & 0.068   \\  
20230  & \zp & 18.482 & 0.032   \\  
21121  & \zp & 18.665 & 0.038   \\  
21929  & \zp & 18.694 & 0.036   \\  
22731  & \zp & 18.686 & 0.051   \\  
23549  & \zp & 18.854 & 0.068   \\  
24357  & \zp & 18.793 & 0.055   \\  
25158  & \zp & 18.974 & 0.072   \\  
25966  & \zp & 19.092 & 0.123   \\  
26758  & \zp & 18.988 & 0.058   \\  
27561  & \zp & 19.03 & 0.076   \\  
28374  & \zp & 19.228 & 0.08   \\  
29187  & \zp & 19.097 & 0.058   \\  
30094  & \zp & 19.281 & 0.06   \\  
30927  & \zp & 19.446 & 0.105   \\  
31730  & \zp & 19.322 & 0.066   \\  
32532  & \zp & 19.304 & 0.063   \\  
33329  & \zp & 19.282 & 0.061   \\  
34132  & \zp & 19.56 & 0.134   \\  
34945  & \zp & 19.646 & 0.124   \\  
35737  & \zp & 19.594 & 0.124   \\  
36534  & \zp & 19.626 & 0.086   \\  
37327  & \zp & 19.708 & 0.15   \\  
38119  & \zp & 19.922 & 0.102   \\  
38985  & \zp & 19.667 & 0.109   \\  
39777  & \zp & 19.745 & 0.098   \\  
40570  & \zp & 19.645 & 0.117   \\  
\enddata 
\label{tab:tab8}
\end{deluxetable}

\begin{deluxetable}{lccc} 
\tablecolumns{4} 
\tablewidth{0pc}
\tablecaption{Photometric observations for GRB 110205A} 
\tabletypesize{\footnotesize}
\tablehead{ 
\colhead{$T-T_0$}  &   \colhead{Filter} &  \colhead{Magnitude$^a$} &
\colhead{1$\sigma$ Error}\\
\,\,\,\,(s) &&}
\startdata 
\multicolumn{4}{c}{PAIRITEL }\\
\cline{1-4}\\
11677  & $J$ & 16.416 & 0.106   \\  
12004  & $J$ & 16.514 & 0.114   \\  
12333  & $J$ & 16.540 & 0.120   \\  
12897  & $J$ & 16.695 & 0.085   \\  
13695  & $J$ & 16.841 & 0.097   \\  
14513  & $J$ & 16.893 & 0.100   \\  
15332  & $J$ & 16.946 & 0.104   \\  
16170  & $J$ & 16.834 & 0.096   \\  
16969  & $J$ & 16.929 & 0.102   \\  
18407  & $J$ & 17.101 & 0.081   \\  
20464  & $J$ & 17.267 & 0.096   \\  
22523  & $J$ & 17.285 & 0.090   \\  
24580  & $J$ & 17.558 & 0.114   \\  
28290  & $J$ & 17.813 & 0.109   \\  
31929  & $J$ & 17.852 & 0.179   \\  
36517  & $J$ & 18.230 & 0.167   \\  
\enddata 
\tablenotetext{a}{PAIRITEL Magnitude are in Vega system.}
\label{tab:tab9}
\end{deluxetable}

\begin{deluxetable}{lccc} 
\tablecolumns{4} 
\tablewidth{0pc}
\tablecaption{Photometric observations for GRB 110205A} 
\tabletypesize{\footnotesize}
\tablehead{ 
\colhead{$T-T_0$}  &   \colhead{Filter} &  \colhead{Magnitude$^a$} &
\colhead{1$\sigma$ Error}\\
\,\,\,\,(s) &&}
\startdata 
\multicolumn{4}{c}{PAIRITEL }\\
\cline{1-4}\\
11677  & $H$ & 15.772 & 0.139   \\  
12004  & $H$ & 15.812 & 0.145   \\  
12333  & $H$ & 15.767 & 0.139   \\  
12897  & $H$ & 15.872 & 0.097   \\  
13695  & $H$ & 15.812 & 0.095   \\  
14513  & $H$ & 16.162 & 0.127   \\  
15332  & $H$ & 16.013 & 0.107   \\  
16170  & $H$ & 16.306 & 0.138   \\  
16969  & $H$ & 16.228 & 0.129   \\  
18407  & $H$ & 16.340 & 0.095   \\  
20464  & $H$ & 16.363 & 0.096   \\  
22523  & $H$ & 16.644 & 0.112   \\  
24580  & $H$ & 16.800 & 0.135   \\  
28290  & $H$ & 17.093 & 0.138   \\  
31929  & $H$ & 17.225 & 0.277   \\  
36517  & $H$ & 17.723 & 0.258   \\  
\enddata 
\tablenotetext{a}{PAIRITEL Magnitude are in Vega system.}
\label{tab:tab10}
\end{deluxetable}

\begin{deluxetable}{lccc} 
\tablecolumns{4} 
\tablewidth{0pc}
\tablecaption{Photometric observations for GRB 110205A} 
\tabletypesize{\footnotesize}
\tablehead{ 
\colhead{$T-T_0$}  &   \colhead{Filter} &  \colhead{Magnitude$^a$} &
\colhead{1$\sigma$ Error}\\
\,\,\,\,(s) &&}
\startdata 
\multicolumn{4}{c}{PAIRITEL }\\
\cline{1-4}\\
11677  & $K$ & 15.127 & 0.183   \\  
12004  & $K$ & 15.228 & 0.205   \\  
12333  & $K$ & 14.816 & 0.143   \\  
12897  & $K$ & 15.345 & 0.149   \\  
13695  & $K$ & 15.224 & 0.132   \\  
14513  & $K$ & 15.352 & 0.152   \\  
15332  & $K$ & 15.360 & 0.142   \\  
16170  & $K$ & 15.480 & 0.161   \\  
16969  & $K$ & 15.359 & 0.146   \\  
18407  & $K$ & 15.478 & 0.107   \\  
20464  & $K$ & 15.772 & 0.140   \\  
22523  & $K$ & 15.967 & 0.155   \\  
24580  & $K$ & 16.103 & 0.178   \\  
28290  & $K$ & 16.255 & 0.140   \\  
31929  & $K$ & 16.586 & 0.300   \\  
36517  & $K$ & 16.782 & 0.210   \\  
\enddata 
\tablenotetext{a}{PAIRITEL Magnitude are in Vega system.}
\label{tab:tab11}
\end{deluxetable}

\begin{deluxetable}{lccc} 
\tablecolumns{4} 
\tablewidth{0pc}
\tablecaption{Photometric observations for GRB 110213A} 
\tabletypesize{\footnotesize}
\tablehead{ 
\colhead{$T-T_0$}  &   \colhead{Filter} &  \colhead{Magnitude$^a$} &
\colhead{1$\sigma$ Error}\\
\,\,\,\,(s) && }
\startdata 
\multicolumn{4}{c}{\swift-UVOT}\\
\cline{1-4}\\
322  & $u$ & 15.27 & 0.07  \\  
342  & $u$ & 15.44 & 0.08   \\  
362  & $u$ & 15.59 & 0.09   \\  
382  & $u$ & 15.66 & 0.09   \\  
402  & $u$ & 15.52 & 0.08   \\  
422  & $u$ & 15.55 & 0.08   \\  
437  & $u$ & 15.58 & 0.03   \\ 
442  & $u$ & 15.62 & 0.09   \\  
462  & $u$ & 15.65 & 0.09   \\  
482  & $u$ & 15.60 & 0.08   \\  
502  & $u$ & 15.60 & 0.09   \\  
522  & $u$ & 15.67 & 0.09   \\  
542  & $u$ & 15.65 & 0.09   \\  
562  & $u$ & 15.76 & 0.13   \\  
722  & $u$ & 15.93 & 0.11   \\  
726  & $u$ & 15.96 & 0.10   \\  
742  & $u$ & 16.07 & 0.25   \\ 
1130  & $u$ & 16.53 & 0.13   \\  
1305  & $u$ & 16.43 & 0.12   \\  
1478  & $u$ & 16.85 & 0.15   \\  
1651  & $u$ & 16.65 & 0.14   \\  
1825  & $u$ & 16.42 & 0.12   \\  
1998  & $u$ & 16.62 & 0.14   \\  
2171  & $u$ & 16.69 & 0.14   \\  
2346  & $u$ & 16.54 & 0.13   \\  
2519  & $u$ & 16.51 & 0.13   \\  
5960  & $u$ & 16.53 & 0.05   \\  
7397  & $u$ & 16.81 & 0.05   \\  
12699  & $u$ & 17.34 & 0.06   \\  
13002  & $u$ & 17.39 & 0.06   \\  
13306  & $u$ & 17.50 & 0.06   \\  
19468  & $u$ & 18.42 & 0.11   \\  
19771  & $u$ & 18.37 & 0.11   \\  
19973  & $u$ & 18.44 & 0.21   \\  
30140  & $u$ & 19.22 & 0.22   \\  
30443  & $u$ & 19.33 & 0.23   \\  
30747  & $u$ & 19.06 & 0.19   \\  
37052  & $u$ & 19.26 & 0.23   \\  
37282  & $u$ & 19.28 & 0.32   \\  
48340  & $u$ & 19.68 & 0.32   \\  
48856  & $u$ & 19.20 & 0.33   \\ 
182877& $u$ & $>20.87$ &    \\ 
\enddata 
\tablenotetext{a}{UVOT magnitude are in the natural UVOT photometric system (Vega). See \cite{Breeveld:2011kx} for the most updated zeropoint and conversion factors.}
\label{tab:tab12}
\end{deluxetable}

\begin{deluxetable}{lccc} 
\tablecolumns{4} 
\tablewidth{0pc}
\tablecaption{Photometric observations for GRB 110213A} 
\tabletypesize{\footnotesize}
\tablehead{ 
\colhead{$T-T_0$}  &   \colhead{Filter} &  \colhead{Magnitude$^a$} &
\colhead{1$\sigma$ Error}\\
\,\,\,\,(s) &&}
\startdata 
\multicolumn{4}{c}{\swift-UVOT}\\
\cline{1-4}\\
577  & $b$ & 16.07 & 0.08  \\  
751  & $b$ & 16.49 & 0.10   \\  
1155  & $b$ & 16.70 & 0.10   \\  
1330  & $b$ & 16.89 & 0.11   \\  
1502  & $b$ & 16.99 & 0.12   \\  
1675  & $b$ & 17.03 & 0.13   \\  
1850  & $b$ & 17.12 & 0.14   \\  
2023  & $b$ & 17.23 & 0.14   \\  
2196  & $b$ & 17.24 & 0.15   \\  
2371  & $b$ & 16.95 & 0.12   \\  
2544  & $b$ & 17.22 & 0.15   \\  
6167  & $b$ & 16.86 & 0.04   \\  
7603  & $b$ & 17.16 & 0.04   \\  
13611  & $b$ & 17.86 & 0.06   \\  
13915  & $b$ & 17.86 & 0.06   \\  
14155  & $b$ & 18.03 & 0.08   \\  
31052  & $b$ & 19.58 & 0.22   \\
431692& $b$ & $>20.91$& 0.22   \\  
\enddata 
\tablenotetext{a}{UVOT magnitude are in the natural UVOT photometric system (Vega). See \cite{Breeveld:2011kx} for the most updated zeropoint and conversion factors.}

\label{tab:tab13}
\end{deluxetable}

\begin{deluxetable}{lccc} 
\tablecolumns{4} 
\tablewidth{0pc}
\tablecaption{Photometric observations for GRB 110213A} 
\tabletypesize{\footnotesize}
\tablehead{ 
\colhead{$T-T_0$}  &   \colhead{Filter} &  \colhead{Magnitude$^a$} &
\colhead{1$\sigma$ Error}\\
\,\,\,\,(s) &&}
\startdata 
\multicolumn{4}{c}{\swift-UVOT}\\
\cline{1-4}\\
652  & $v$ & 15.66 & 0.11  \\  
825  & $v$ & 16.09 & 0.14   \\  
1056  & $v$ & 16.15 & 0.15   \\  
1231  & $v$ & 15.99 & 0.13   \\  
1404  & $v$ & 15.90 & 0.12   \\  
1577  & $v$ & 16.36 & 0.16   \\  
1751  & $v$ & 16.27 & 0.15   \\  
1925  & $v$ & 16.40 & 0.18   \\  
2098  & $v$ & 16.64 & 0.20   \\  
2272  & $v$ & 16.85 & 0.23   \\  
2445  & $v$ & 16.32 & 0.15   \\  
2618  & $v$ & 16.46 & 0.17   \\  
6782  & $v$ & 16.32 & 0.05   \\  
8218  & $v$ & 16.55 & 0.06   \\  
24339  & $v$ & 18.48 & 0.20   \\  
24644  & $v$ & 18.64 & 0.22   \\  
24948  & $v$ & 18.29 & 0.16   \\  
42061  & $v$ & 19.18 & 0.36   \\  
451502&$v$ & $>20.66$&\\
\enddata 
\tablenotetext{a}{UVOT magnitude are in the natural UVOT photometric system (Vega). See \cite{Breeveld:2011kx} for the most updated zeropoint and conversion factors.}

\label{tab:tab14}
\end{deluxetable}

\begin{deluxetable}{lccc} 
\tablecolumns{4} 
\tablewidth{0pc}
\tablecaption{Photometric observations for GRB 110213A} 
\tabletypesize{\footnotesize}
\tablehead{ 
\colhead{$T-T_0$}  &   \colhead{Filter} &  \colhead{Magnitude$^a$} &
\colhead{1$\sigma$ Error}\\
\,\,\,\,(s) &&}
\startdata 
\multicolumn{4}{c}{\swift-UVOT}\\
\cline{1-4}\\
104  & $white$ & 17.33 & 0.20   \\  
112  & $white$ & 17.12 & 0.18   \\  
120  & $white$ & 17.51 & 0.23   \\  
128  & $white$ & 16.76 & 0.13   \\  
136  & $white$ & 16.67 & 0.13   \\  
144  & $white$ & 16.62 & 0.12   \\  
152  & $white$ & 16.07 & 0.09   \\  
160  & $white$ & 16.42 & 0.11   \\  
168  & $white$ & 16.19 & 0.09   \\  
176  & $white$ & 16.04 & 0.08   \\  
184  & $white$ & 16.30 & 0.10   \\  
192  & $white$ & 16.01 & 0.08   \\  
200  & $white$ & 16.08 & 0.09   \\  
208  & $white$ & 16.03 & 0.08   \\  
216  & $white$ & 15.93 & 0.08   \\  
224  & $white$ & 15.99 & 0.08   \\  
232  & $white$ & 15.98 & 0.08   \\  
240  & $white$ & 15.83 & 0.07   \\  
248  & $white$ & 15.83 & 0.09   \\  
592  & $white$ & 16.02 & 0.10   \\  
600  & $white$ & 16.17 & 0.08   \\  
602  & $white$ & 16.09 & 0.05   \\ 
608  & $white$ & 16.03 & 0.07   \\  
768  & $white$ & 16.44 & 0.10   \\  
775  & $white$ & 16.33 & 0.05   \\ 
776  & $white$ & 16.35 & 0.08   \\  
784  & $white$ & 16.15 & 0.09   \\  
941  & $white$ & 16.45 & 0.02   \\  
1176  & $white$ & 16.69 & 0.10   \\  
1179  & $white$ & 16.63 & 0.06   \\  
1184  & $white$ & 16.55 & 0.09   \\  
1352  & $white$ & 16.70 & 0.10   \\  
1354  & $white$ & 16.74 & 0.06   \\ 
1360  & $white$ & 16.81 & 0.10   \\  
1520  & $white$ & 16.73 & 0.10   \\  
1527  & $white$ & 16.83 & 0.07   \\  
1528  & $white$ & 16.77 & 0.10   \\  
1536  & $white$ & 17.04 & 0.15   \\  
1695  & $white$ & 16.84 & 0.09   \\  
1705  & $white$ & 16.84 & 0.09   \\  
1865  & $white$ & 17.02 & 0.15   \\  
1875  & $white$ & 16.87 & 0.09   \\  
1885  & $white$ & 17.54 & 0.22   \\  
2035  & $white$ & 16.59 & 0.17   \\  
2045  & $white$ & 16.95 & 0.10   \\  
2055  & $white$ & 16.98 & 0.12   \\  
2215  & $white$ & 16.87 & 0.10   \\  
2225  & $white$ & 16.84 & 0.09   \\  
2385  & $white$ & 16.92 & 0.14   \\  
2395  & $white$ & 16.94 & 0.10   \\  
2405  & $white$ & 16.89 & 0.14   \\  
2555  & $white$ & 17.18 & 0.30   \\  
2565  & $white$ & 16.97 & 0.10   \\  
2575  & $white$ & 16.78 & 0.10   \\  
6371  & $white$ & 16.83 & 0.03   \\  
7807  & $white$ & 17.04 & 0.03   \\  
183368  & $white$ & 21.57 & 0.31   \\  
\enddata 
\tablenotetext{a}{UVOT magnitude are in the natural UVOT photometric system (Vega). See \cite{Breeveld:2011kx} for the most updated zeropoint and conversion factors.}

\label{tab:tab4}
\end{deluxetable}

\begin{deluxetable}{lccc} 
\tablecolumns{4} 
\tablewidth{0pc}
\tablecaption{Photometric observations for GRB 110213A} 
\tabletypesize{\footnotesize}
\tablehead{ 
\colhead{$T-T_0$}  &   \colhead{Filter} &  \colhead{Magnitude} &
\colhead{1$\sigma$ Error}\\
\,\,\,\,(s) &&(AB)}
\startdata 
\multicolumn{4}{c}{KAIT}\\
\cline{1-4}\\
74  & $clear$ & 17.34 & 0.10   \\  
77  & $clear$ & 17.03 & 0.08   \\  
81  & $clear$ & 17.05 & 0.08   \\  
84  & $clear$ & 16.91 & 0.07   \\  
87  & $clear$ & 16.88 & 0.06   \\  
90  & $clear$ & 16.68 & 0.06   \\  
93  & $clear$ & 16.56 & 0.05   \\  
97  & $clear$ & 16.59 & 0.05   \\  
100  & $clear$ & 16.47 & 0.05   \\  
103  & $clear$ & 16.37 & 0.05   \\  
195  & $clear$ & 15.07 & 0.01   \\  
295  & $clear$ & 14.96 & 0.01   \\  
395  & $clear$ & 15.04 & 0.01   \\  
495  & $clear$ & 15.22 & 0.01   \\  
595  & $clear$ & 15.34 & 0.01   \\  
695  & $clear$ & 15.47 & 0.01   \\  
795  & $clear$ & 15.57 & 0.01   \\  
894  & $clear$ & 15.67 & 0.01   \\  
992  & $clear$ & 15.75 & 0.01   \\  
1092  & $clear$ & 15.84 & 0.01   \\  
1192  & $clear$ & 15.90 & 0.01   \\  
1290  & $clear$ & 15.97 & 0.01   \\  
1390  & $clear$ & 16.03 & 0.01   \\  
1488  & $clear$ & 16.07 & 0.01   \\  
1588  & $clear$ & 16.12 & 0.01   \\  
1655  & $clear$ & 16.13 & 0.01   \\  
1722  & $clear$ & 16.15 & 0.01   \\  
1788  & $clear$ & 16.19 & 0.01   \\  
1853  & $clear$ & 16.21 & 0.01   \\  
1920  & $clear$ & 16.25 & 0.01   \\  
1986  & $clear$ & 16.27 & 0.01   \\  
2053  & $clear$ & 16.33 & 0.01   \\  
2120  & $clear$ & 16.27 & 0.01   \\  
2186  & $clear$ & 16.24 & 0.01   \\  
2253  & $clear$ & 16.26 & 0.01   \\  
2320  & $clear$ & 16.27 & 0.01   \\  
2386  & $clear$ & 16.14 & 0.01   \\  
2454  & $clear$ & 16.12 & 0.01   \\  
2518  & $clear$ & 16.16 & 0.01   \\  
2585  & $clear$ & 16.13 & 0.01   \\  
78843  & $clear$ & 19.85 & 0.10   \\  
84641  & $clear$ & 20.22 & 0.10   \\  
86515  & $clear$ & 20.14 & 0.15   \\  
\enddata 
\label{tab:tab15}
\end{deluxetable}

\begin{deluxetable}{lccc} 
\tablecolumns{4} 
\tablewidth{0pc}
\tablecaption{Photometric observations for GRB 110213A} 
\tabletypesize{\footnotesize}
\tablehead{ 
\colhead{$T-T_0$}  &   \colhead{Filter} &  \colhead{Magnitude} &
\colhead{1$\sigma$ Error}\\
\,\,\,\,(s) &&}
\startdata 
\multicolumn{4}{c}{Palomar telescope}\\
\cline{1-4}\\
2304  & \gp & 16.449 & 0.009   \\  
2892  & \gp & 16.364 & 0.005   \\  
3480  & \gp & 16.260 & 0.006   \\  
3775  & \gp & 16.359 & 0.007   \\  
4343  & \gp & 16.195 & 0.007   \\  
4900  & \gp & 16.072 & 0.005   \\  
5463  & \gp & 16.184 & 0.006   \\  
6026  & \gp & 16.190 & 0.006   \\  
6579  & \gp & 16.296 & 0.006   \\  
88212  & \gp & 20.669 & 0.155   \\ 
\enddata 
\label{tab:tab16}
\end{deluxetable}

\begin{deluxetable}{lccc} 
\tablecolumns{4} 
\tablewidth{0pc}
\tablecaption{Photometric observations for GRB 110213A} 
\tabletypesize{\footnotesize}
\tablehead{ 
\colhead{$T-T_0$}  &   \colhead{Filter} &  \colhead{Magnitude} &
\colhead{1$\sigma$ Error}\\
\,\,\,\,(s) &(AB)&}
\startdata 
\multicolumn{4}{c}{KAIT }\\
\cline{1-4}\\
128  & $V$ & 15.81 & 0.03   \\  
228  & $V$ & 15.13 & 0.02   \\  
328  & $V$ & 15.14 & 0.02   \\  
428  & $V$ & 15.24 & 0.02  \\  
528  & $V$ & 15.36 & 0.02   \\  
628  & $V$ & 15.55 & 0.02   \\  
728  & $V$ & 15.67 & 0.03   \\  
828  & $V$ & 15.675 & 0.03   \\  
926  & $V$ & 15.84 & 0.03   \\  
1026  & $V$ & 15.93 & 0.03   \\  
1126  & $V$ & 15.97 & 0.03   \\  
1226  & $V$ & 16.05 & 0.03   \\  
1324  & $V$ & 16.08 & 0.03   \\  
1422  & $V$ & 16.13 & 0.03   \\  
1522  & $V$ & 16.19 & 0.03   \\  
\enddata 
\label{tab:tab17}
\end{deluxetable}

\begin{deluxetable}{lccc} 
\tablecolumns{4} 
\tablewidth{0pc}
\tablecaption{Photometric observations for GRB 110213A} 
\tabletypesize{\footnotesize}
\tablehead{ 
\colhead{$T-T_0$}  &   \colhead{Filter} &  \colhead{Magnitude} &
\colhead{1$\sigma$ Error}\\
\,\,\,\,(s) &&}
\startdata 
\multicolumn{4}{c}{Palomar Telescope}\\
\cline{1-4}\\
162  & \rp & 15.326 & 0.006   \\  
248  & \rp & 14.956 & 0.005   \\  
582  & \rp & 15.271 & 0.005   \\  
839  & \rp & 15.584 & 0.003   \\  
1131  & \rp & 15.853 & 0.018   \\  
1247  & \rp & 15.912 & 0.005   \\  
1575  & \rp & 16.094 & 0.006   \\  
1862  & \rp & 16.198 & 0.005   \\  
2450  & \rp & 16.103 & 0.005   \\  
3038  & \rp & 16.121 & 0.005   \\  
3632  & \rp & 15.976 & 0.006   \\  
4205  & \rp & 15.974 & 0.005   \\  
4762  & \rp & 15.802 & 0.005   \\  
5315  & \rp & 15.804 & 0.005   \\  
5888  & \rp & 15.886 & 0.005   \\  
6441  & \rp & 15.937 & 0.006   \\  
6994  & \rp & 16.083 & 0.006   \\  
88628  & \rp & 20.303 & 0.083   \\  
\cline{1-4}\\
\multicolumn{4}{c}{Gemini-N Telescope}\\
\cline{1-4}\\
526800 & \rp & 24.27 & 0.13 \\
\enddata 
\label{tab:tab18}
\end{deluxetable}

\begin{deluxetable}{lccc} 
\tablecolumns{4} 
\tablewidth{0pc}
\tablecaption{Photometric observations for GRB 110213A} 
\tabletypesize{\footnotesize}
\tablehead{ 
\colhead{$T-T_0$}  &   \colhead{Filter} &  \colhead{Magnitude} &
\colhead{1$\sigma$ Error}\\
\,\,\,\,(s) &&(AB)}
\startdata 
\multicolumn{4}{c}{KAIT telescope}\\
\cline{1-4}\\
161  & $I$ & 14.64 & 0.02   \\  
261  & $I$ & 14.37 & 0.02   \\  
361  & $I$ & 14.41 & 0.02   \\  
461  & $I$ & 14.60 & 0.02   \\  
561  & $I$ & 14.74 & 0.02   \\  
661  & $I$ & 14.84 & 0.02   \\  
761  & $I$ & 14.97 & 0.02   \\  
861  & $I$ & 15.05 & 0.02   \\  
959  & $I$ & 15.16 & 0.02   \\  
1059  & $I$ & 15.24 & 0.02   \\  
1159  & $I$ & 15.32& 0.02   \\  
1257  & $I$ & 15.39 & 0.02   \\  
1357  & $I$ & 15.42 & 0.02   \\  
1455  & $I$ & 15.51 & 0.02   \\  
1555  & $I$ & 15.54 & 0.03   \\  
1622  & $I$ & 15.56 & 0.03   \\  
1688  & $I$ & 15.63 & 0.03   \\  
1755  & $I$ & 15.64 & 0.03   \\  
1822  & $I$ & 15.63 & 0.03   \\  
1886  & $I$ & 15.64 & 0.03   \\  
1953  & $I$ & 15.68 & 0.03   \\  
2020  & $I$ & 15.72 & 0.03   \\  
2086  & $I$ & 15.76 & 0.03   \\  
2153  & $I$ & 15.69 & 0.03   \\  
2220  & $I$ & 15.64 & 0.03   \\  
2287  & $I$ & 15.67 & 0.03   \\  
2353  & $I$ & 15.63 & 0.03   \\  
2420  & $I$ & 15.58 & 0.03   \\  
2485  & $I$ & 15.57 & 0.03   \\  
2551  & $I$ & 15.54 & 0.03   \\  
2618  & $I$ & 15.53 & 0.03   \\  
\multicolumn{4}{c}{Palomar telescope}\\
\cline{1-4}\\
334  & \ip & 14.322 & 0.006   \\  
667  & \ip & 14.843 & 0.005   \\  
924  & \ip & 15.103 & 0.005   \\  
1343  & \ip & 15.407 & 0.005   \\  
1661  & \ip & 15.573 & 0.005   \\  
2013  & \ip & 15.673 & 0.003   \\  
2601  & \ip & 15.528 & 0.005   \\  
3183  & \ip & 15.550 & 0.003   \\  
3913  & \ip & 15.439 & 0.006   \\  
4481  & \ip & 15.183 & 0.003   \\  
5039  & \ip & 15.124 & 0.005   \\  
5601  & \ip & 15.265 & 0.005   \\  
6164  & \ip & 15.258 & 0.005   \\  
6717  & \ip & 15.414 & 0.006   \\  
89048  & \ip & 19.611 & 0.054   \\  
262255  & \ip & 22.315 & 0.371   \\  
\multicolumn{4}{c}{Gemini telescope}\\
\cline{1-4}\\
526800 &\ip & 23.68 & 0.09 \\
\enddata 
\label{tab:tab19}
\end{deluxetable}

\begin{deluxetable}{lccc} 
\tablecolumns{4} 
\tablewidth{0pc}
\tablecaption{Photometric observations for GRB 110213A} 
\tabletypesize{\footnotesize}
\tablehead{ 
\colhead{$T-T_0$}  &   \colhead{Filter} &  \colhead{Magnitude} &
\colhead{1$\sigma$ Error}\\
\,\,\,\,(s) &&}
\startdata 
\multicolumn{4}{c}{Palomar telescope}\\
\cline{1-4}\\
419  & \zp & 13.910 & 0.004   \\  
753  & \zp & 14.383 & 0.006   \\  
1010  & \zp & 14.630 & 0.007   \\  
1433  & \zp & 14.928 & 0.010   \\  
1747  & \zp & 15.076 & 0.011   \\  
2159  & \zp & 15.121 & 0.008   \\  
2746  & \zp & 14.877 & 0.007   \\  
3334  & \zp & 14.793 & 0.006   \\  
4056  & \zp & 14.843 & 0.006   \\  
4624  & \zp & 14.679 & 0.005   \\  
5177  & \zp & 14.604 & 0.005   \\  
5744  & \zp & 14.721 & 0.006   \\  
6302  & \zp & 14.762 & 0.006   \\  
6855  & \zp & 14.929 & 0.006   \\ \enddata 
\label{tab:tab20}
\end{deluxetable}

\end{document}